\documentclass[12pt,preprint,number,sort&compress,float]{elsarticle}
\usepackage{amsmath,amssymb,bm,bbm}
\usepackage{graphicx}
\usepackage{subcaption}
\usepackage{color}
\usepackage[dvipsnames]{xcolor}
\usepackage[papersize={8.5in,11in}]{geometry}
\usepackage[colorlinks=true]{hyperref}
\usepackage[mathscr]{euscript}
\usepackage[section]{placeins}
\usepackage{comment}
\hypersetup{
    bookmarks=true,         
    unicode=false,          
    pdftoolbar=true,        
    pdfmenubar=true,        
    pdffitwindow=false,     
    pdfstartview={FitH},    
    pdfsubject={},   
    pdfcreator={},   
    pdfproducer={}, 
    pdfkeywords={} {} {}, 
    pdfnewwindow=true,      
    colorlinks=true,       
    linkcolor=magenta, 
    citecolor=blue,        
    filecolor=magenta,      
    urlcolor=blue           
}

\usepackage{tikz}
\usepackage{tikz-cd}
\usetikzlibrary{arrows}
\usetikzlibrary{intersections}
\usetikzlibrary{shapes.geometric}
\usetikzlibrary{decorations.pathmorphing, patterns,shapes}
\usetikzlibrary{decorations.markings}

\tikzset{
  mid arrow/.style={postaction={decorate,decoration={
        markings,
        mark=at position .575 with {\arrow[#1]{stealth}}
      }}},
  near arrow/.style={postaction={decorate,decoration={
        markings,
        mark=at position .275 with {\arrow[#1]{Stealth}}
      }}},
   far arrow/.style={postaction={decorate,decoration={
        markings,
        mark=at position .800 with {\arrow[#1]{stealth}}
      }}},
}

\usepackage{amsfonts}
\usepackage{upgreek}
\usepackage{slashed}
\usepackage{latexsym}

\newcommand{\beq}{\begin{equation}}
\newcommand{\eeq}{\end{equation}}
\def\bea{\begin{eqnarray}}
\def\eea{\end{eqnarray}}
\newcommand{\nn}{\nonumber }

\renewcommand{\approx}{\simeq}

\renewcommand{\Im}{\text{Im}}

\renewcommand{\tilde}{\widetilde}

\renewcommand{\geq}{\geqslant}

\definecolor{wrongultramarine}{rgb}{1,0.5,0}

\newcommand{\rd}{{\rm d}}
\newcommand{\sgn}{{\rm sgn\,}}
\newcommand{\Tr}{{\rm \, Tr\,}}
\newcommand{\ce}{\mathcal{E}}

\newcommand{\onlinecite}[1]{\cite{#1}}

\begin{document}

\title{Linear in temperature resistivity in the limit of zero temperature\\
from the time reparameterization soft mode}

\author{Haoyu Guo\corref{cor1}}

\author{Yingfei Gu}

\author{Subir Sachdev}
\address{Department of Physics, Harvard University, Cambridge MA 02138, USA}

\cortext[cor1]{Corresponding Author}

\date{\today
\\
\vspace{0.4in}}

\begin{abstract}
The most puzzling aspect of the `strange metal' behavior of correlated electron compounds is that the linear in temperature resistivity often extends down to low temperatures, lower than natural microscopic energy scales. We consider recently proposed deconfined critical points (or phases) in models of electrons in large dimension lattices with random nearest-neighbor exchange interactions. The criticality is in the class of Sachdev-Ye-Kitaev models, and exhibits a time reparameterization soft mode representing gravity in dual holographic theories. We compute the low temperature resistivity in a large $M$ limit of models with SU($M$) spin symmetry, and find that the dominant temperature dependence arises from this soft mode. The resistivity is  
linear in temperature down to zero temperature at the critical point, with a co-efficient universally proportional to the product of the residual resistivity and the co-efficient of the linear in temperature specific heat. We argue that the time reparameterization soft mode offers a promising and generic mechanism for resolving the strange metal puzzle.
\end{abstract}

\maketitle
\begin{center}
\vspace{1in}
    \href{https://arxiv.org/abs/2004.05182}{arXiv:2004.05182}
\end{center}
\newpage
\tableofcontents

\section{Introduction}
\label{sec:intro}

The strange metal problem has been a central puzzle in the theory of correlated electron systems since its identification in the `normal' state of the cuprate high temperature superconductors \cite{Takagi92,Taillefer10,MacKenzie13,Greene17,Bakr19,Legros18,Pablo19,Paglione19}. While the anomalous behavior is often identified by a linear temperature ($T$) dependence in the resistivity, it is important to distinguish different $T$ regimes. At very high $T$, the resistivity can usually be understood via a resummed high $T$ expansion of electron-electron and electron-phonon interactions \cite{Bakr19,Huse06,Georges13,Werman1,Werman2}; we shall not be interested in this regime in the present paper. At lower $T$, below the electron-electron interaction and Debye energy scales, the linear-in-$T$ resistivity has been described by various lattice models \cite{PG98,Zhang2017,Balents17,Patel2017,DC18,Guo2019,PatelSS19,PatelKim2019} exhibiting a non-Fermi liquid regime with criticality similar to that of the Sachdev-Ye-Kitaev (SYK) model \cite{SY92,kitaev2015talk}.  
But in these models, the linear-in-$T$ resistivity holds only down to a `coherence' temperature below which quasiparticles emerge and the resistivity is Fermi liquid-like. It is possible to consider models with  `resonant' interactions to suppress the coherence temperature \cite{PatelSS19}, but generically non-resonant interactions are present, and will provide a low $T$ cutoff to the linear-in-$T$ resistivity. So experimental observations of linear $T$ resistivity at the lowest $T$ 
\cite{Greene17,Legros18,Pablo19,Paglione19} are not fully understood by these models.

A significant motivation for our work was provided by the recent 
numerical study by Cha {\it et al.} \cite{Cha19} of a non-random Hubbard model on a large dimension lattice, with additional random nearest-neighbor exchange interactions with zero mean. They found a continuous metal-insulator transition at half-filling, with spin correlations exhibiting SYK behavior at the critical point. Moreover, the critical point resistivity was linear in $T$ down to the lowest numerically accessible $T$.

This paper will present an analysis of the electrical transport properties of a $t$-$J$ model that was originally introduced by Parcollet and Georges \cite{PG98}. This model has a non-random nearest-neighbor hopping on a large dimension lattice, and random nearest-neighbor exchange interactions with zero mean.
We will examine this $t$-$J$ model at the deconfined critical point at a non-zero doping $p=p_c$ that was found recently by 
Joshi {\it et al.} \cite{Joshi2019}. In a large $M$ analysis (in a model with SU($M$) spin symmetry), we find a linear-in-$T$ resistivity as $T \rightarrow 0$, above a background residual resistivity (see (\ref{rhoT}) below).
By considering a particular particle-hole symmetric limit of our analysis, we will also obtain similar results at a large $M$ deconfined metal-insulator critical point found by Tarnopolsky { \it et al.} \cite{Grisha2020} in a Hubbard model, which then provides a potential understanding of the linear-in-$T$ resistivity observed numerically by Cha {\it et al.} \cite{Cha19}. In both cases, the deconfined critical point is accompanied by an adjoining deconfined critical phase on the side of the disordered Fermi liquid \cite{Joshi2019,Grisha2020,Fu18}, at least in the large $M$ limit; we expect very similar results to apply to these critical phases, but will not describe them explicitly.

As an aside, we briefly review the past theoretical work on the models mentioned above. Parcollet and Georges originally \cite{PG98} examined their model in a different $M$ limit, and the differences are reviewed in \ref{app:PG}. They found a linear-in-$T$ resistivity only above a non-zero coherence temperature (as noted above). Joshi {\it et al.} \cite{Joshi2019} reconsidered the same $t$-$J$ model in a large $M$ limit related to that introduced by Fu {\it et al.} \cite{Fu18}, and also presented a renormalization group analysis which determined key exponents to all loop order. From these analyses, Joshi {\it et al.} \cite{Joshi2019} argued that there was a deconfined critical point at a non-zero doping $p=p_c$, or possibly a critical phase over a finite range of $p$, exhibiting SYK criticality in spin correlations down to $T=0$. Tarnopolsky { \it et al.} \cite{Grisha2020} have applied the large $M$ limit of Joshi {\it et al.} \cite{Joshi2019} (see also Refs.~\cite{Florens02,Haule02,Haule03,Florens04,Florens13}) to the Hubbard model, and argued that the corresponding deconfined critical point provides an understanding of the electron and spin spectral functions at the metal-insulator transition in the numerical study of Cha {\it et al.} \cite{Cha19}.

We will begin by describing the lattice $t$-$J$ model, its large dimension limit, and the subsequent large $M$ limit in Section~\ref{sec:lattice}. These limits lead to saddle point equations for Green's functions of fermionic spinons $G_f$ and bosonic holons $G_b$. These equations have critical SYK-like solutions \cite{Fu18,Joshi2019} for $G_f$ and $G_b$ with spinon scaling dimension $\Delta_f$, and holon scaling dimension $\Delta_b$, which obey $\Delta_f + \Delta_b =1/2$. We shall only consider the case identified as the deconfined critical point by Joshi {\it et al.} \cite{Joshi2019} with $\Delta_f = \Delta_b = 1/4$, which yields exponents in agreement with their renormalization group analysis. The large $M$ limit also has a phase with $1/4 < \Delta_f < 1/2$ (and $\Delta_b = 1/2-\Delta_f$) which we will not consider, but for which we expect very similar transport properties.

We compute the leading `conformal' contribution to the electrical conductivity in Section~\ref{sec:conductivity}. Because of the exponent identity $\Delta_f + \Delta_b = 1/2$, the electron Green's function $G_c$, which is the product of $G_f$ and $G_b$ in the time domain, has a $1/\tau$ decay in imaginary time, $\tau$. This has an important consequence for the $T\rightarrow 0$ limit of the resistivity at the deconfined critical point: the leading low $T$ form of the Green's functions is specified by a $R$) conformal invariance, and these yield a temperature independent `residual' resistivity $\rho(0)$. This is in contrast to the large $M$ limit of Parcollet and Georges \cite{PG98} (\ref{app:PG}), and the lattices of SYK islands \cite{Zhang2017,Balents17,Patel2017,DC18,Guo2019,PatelSS19}, where the resistivity is linear-in-$T$ in the intermediate $T$ regime where the conformal solutions apply.

Our main new results appear in Section~\ref{sec:conformalcorrection}-\ref{sec:eval_conductivity}. To obtain any $T$ dependence in the resistivity at the deconfined critical point as $T \rightarrow 0$, we have to consider corrections to the Green's function beyond the conformal form. These have been described in some detail for the SYK model in Refs.~\cite{kitaev2015talk,Maldacena2016,Bagrets2016,kitaev2017,GKST2019}, and we will extend their analyses to the large $M$ deconfined critical points of $t$-$J$ and Hubbard models. The dominant non-conformal corrections to the Green's functions arise from a 
time reparameterization soft mode governed by a Schwarzian action. The boson and fermion Green's functions then have a frequency ($\omega$) and $T$ dependence of the form ($\omega, T \ll J$, $\omega/T$ arbitrary)
\beq
G(\omega) \sim T^{2 \Delta -1} \left[\mathcal{G}_{\rm conformal} (\omega/T) + \alpha_G \, \frac{T}{J} \,\mathcal{G}_{\rm non-conformal} (\omega/T) + \ldots \right]\,.
\label{Gcnc}
\eeq
Here $\Delta$ is the scaling dimension of the field operator, $\mathcal{G}_{\rm conformal}$ and $\mathcal{G}_{\rm non-conformal}$ are universal 
scaling functions of $\omega/T$, which we compute exactly here for the deconfined critical points. The dimensionless number $\alpha_G$ is non-universal and depends upon $t/J$, where $t$ is the electron hopping amplitude and $J$ is the mean-square exchange  (we assume $J$ and $t$ are of the same order). Determination of $\alpha_G$ requires a full numerical solution of the saddle-point equations which will not be performed here. The scaling function $\mathcal{G}_{\rm non-conformal}$ is the contribution of the time reparameterization zero mode, and it has a prefactor which is linear in $T$, which is key to our results. It is important to note that this linearity in $T$ is {\it not \/} the result of a Taylor expansion in integer powers of $T$.
It is instead a non-trivial scaling dimension associated with the manner in which time reparameterization is broken to a conformal SL(2,$R$) symmetry \cite{Maldacena2016,kitaev2017,GKST2019}. The higher order corrections to (\ref{Gcnc}) depend upon transcendental powers of $T$, determined by the scaling dimensions of composite operators \cite{Maldacena2016,kitaev2017}. Finally, we emphasize that the expansion in (\ref{Gcnc}) is carried out {\it at\/} the large dimension and large $M$ saddle point, and the corrections to conformality are suppressed only by powers of $T/J$.

We will compute the leading $T$ dependence of the resistivity, $\rho(T)$, of the deconfined critical point arising from (\ref{Gcnc}) in Section~\ref{sec:eval_conductivity}; we find 
\beq
\rho(T)= \rho(0) \left(1 + 8 \alpha_G \, \frac{T}{J} + \ldots \right)\,.
\label{rhoT}
\eeq
This has the advertised the linear in $T$ dependence as $T \rightarrow 0$, with slope $\zeta \equiv 8\rho(0)\alpha_G/J$, arising from the contribution of the time reparameterization soft mode. For the SYK model, it is known \cite{Maldacena2016,kitaev2017,GKST2019} that $\alpha_G$ is proportional to $\alpha_S$, the dimensionless co-efficient of the Schwarzian term in the effective action, and the latter determines the linear in $T$ co-efficient of the low $T$ specific heat. We find the same relationship here for the deconfined critical point. In particular, the specific heat per site and per spin, $C=\gamma T$,
obeys
\beq
C = 4 \pi^2 \alpha_S \, \frac{T}{J} + \ldots
\eeq
as $T \rightarrow 0$. The ratio $\alpha_G/\alpha_S$ is a universal number, which we will determine exactly for the deconfined critical points of the $t$-$J$ and Hubbard models. Our large $M$ model yields the following universal relation between $\zeta,~\gamma$ and $\rho(0)$,  
\begin{equation}\label{eq:zetagammarelation}
  \frac{\zeta}{\rho(0)\gamma} \equiv \frac{2}{\pi^2}\frac{\alpha_G}{\alpha_S} = g(k,\theta_f,\theta_b),
\end{equation} where $g$ is a universal function specified in (\ref{defgk}) that depends upon $k$ (defined in \eqref{eq:k}; $k=1/2$ for the SU(2) model of interest) and the angles $\theta_{f,b}$ which control the particle-hole asymmetry of the spinons and holons; these angles are determined from
doping $p$ from half filling by (\ref{eq:Ltf},\ref{eq:LtB}).

\section{Lattice Hamiltonian and Schwinger-Dyson equation}
\label{sec:lattice}

We consider a $t$-$J$ model on a high dimensional lattice with random nearest-neighbor $J_{ij}$ as
in Ref.~\onlinecite{PG98}. We need a large $M$ limit to solve the problem, but the nature of the large $M$ limit will be different from that in Ref.~\onlinecite{PG98}.
We follow the approach of Ref.~\onlinecite{Joshi2019}, and generalize the SU(2) spin index
of the electron, $\alpha$, to $\alpha = 1 \ldots M$ a SU($M$) spin index (as in Ref.~\onlinecite{SY92}).
Unlike Ref.~\onlinecite{PG98}, we also introduce an orbital index, $\ell = 1 \ldots M'$, and fractionalize the electron as
\beq
c_{i \ell \alpha} = b_{i \ell}^\dagger f_{i \alpha} \,, \label{defc}
\eeq
where $b_{i \ell}$ is a complex `slave boson' (see~\ref{app:PG} for the alternative fractionalization
of Ref.~\onlinecite{PG98}).
Then we take the large $M$ and $M'$ limit, at fixed
\beq\label{eq:k}
k = \frac{M'}{M}.
\eeq
The representation (\ref{defc}) has a U(1) gauge invariance
\beq
b_{i \ell} \rightarrow b_{i \ell} e^{i \phi_i (\tau)} \quad , \quad f_{i \alpha} \rightarrow f_{i \alpha} e^{i \phi_i (\tau)} \label{u1gauge}
\eeq
We shall be interested in the sector in which the U(1) gauge charge is fixed on each site by
\beq
\sum_{\alpha=1}^M f_{i \alpha}^\dagger f_{i \alpha} + \sum_{\ell=1}^{M'} b_{i \ell}^\dagger b_{i \ell} = \frac{M}{2}\,.
\label{const}
\eeq
Also, the doping density, $p$, will be fixed by a chemical potential
\beq
 \sum_{\ell=1}^{M'} \left\langle b_{i \ell}^\dagger b_{i \ell} \right\rangle  = M' p.
\label{doping}
\eeq

The Hamiltonian we shall study is just the $t$-$J$ model:
\bea
H = \sum_{\langle i j \rangle,\ell,\alpha} t_{ij} \left( c_{i \ell \alpha}^\dagger c_{j \ell \alpha} + \mbox{H.c.} \right) +  \sum_{\langle i j \rangle,\alpha\beta} J_{ij} f_{i \alpha}^\dagger f_{i \beta} f_{j \beta}^\dagger f_{j \alpha} \,,
\eea
along with a chemical potential to impose (\ref{doping}), and a Lagrange multiplier to impose (\ref{const}). The $J_{ij}$ are random variables chosen just as in Ref.~\onlinecite{PG98} with
\beq
J_{ij} = \frac{J}{\sqrt{M z}} \epsilon_{ij}, \quad \overline{\epsilon_{ij}} = 0, \quad
\overline{\epsilon_{ij}^2} = 1\,,
\eeq
where $z$ is the lattice co-ordination number.
The $t_{ij}$ are non-random, but scaled differently with $M$ from Ref.~\onlinecite{PG98}
\beq
t_{ij} = \frac{t}{\sqrt{M z }} \,. \label{t1}
\eeq

\subsection{Large $z$ limit}
\label{sec:largeN}

We follow precisely the procedure in Ref.~\onlinecite{PG98}, and reduce the large $z$ limit to a single-site path integral
\bea
\mathcal{Z} &=& \int \mathcal{D} f_\alpha \mathcal{D} b_\ell \mathcal{D} \lambda
e^{- \mathcal{S}} \nn \\
\mathcal{S} &=& \int_0^{1/T} d \tau \left[ \sum_{\ell} b_{\ell}^\dagger \left( \frac{\partial}{\partial \tau} + i \lambda \right) b_{\ell}
+ \sum_{\alpha} f_{\alpha}^\dagger \left( \frac{\partial}{\partial \tau} + \epsilon_0 + i \lambda \right) f_{\alpha} - i \lambda \frac{M}{2} \right] \nn \\
&~&~+\frac{t^2}{M} \sum_{\ell,\alpha} \int_{0}^{1/T} d \tau d \tau' 
f^\dagger_{\alpha}(\tau)b_{\ell}(\tau) G_c(\tau-\tau') b^\dagger_{\ell}(\tau') f_{\alpha}(\tau')  
\nn \\ &~&~-
\frac{J^2}{2M} \sum_{\alpha,\beta} \int_{0}^{1/T} d \tau d \tau'  Q(\tau-\tau') f_{\alpha}^\dagger (\tau) f_{\beta} (\tau) f_{\beta}^\dagger (\tau') f_{\alpha} (\tau') \,.
\label{L}
\eea
Here $T$ is the temperature, $\lambda$ is the Lagrange multiplier imposing Eq.~(\ref{const}) and $\epsilon_0$ is to be determined to satisfy (\ref{doping}). Note that under the U(1) gauge transformation
(\ref{u1gauge}),
\beq
\lambda \rightarrow \lambda - \partial_\tau \phi\,.
\eeq
As in Ref.~\onlinecite{SY92}, decoupling the large $z$ path integral introduces fields analogous to $G_c$ and $Q$ which are off-diagonal in the SU($M$) and SU($M'$) indices. We have assumed above that the large $z$ limit is dominated by the saddle point in which these fields are SU($M$) and SU($M'$) diagonal. This requires that the large $z$ limit is taken {\it before\/} the large $M$  and $M'$ limits.
In this procedure, $G_c (\tau)$ and $Q(\tau)$ are {\it fixed\/} fields which act as external sources on $\mathcal{Z}$. These are to be determined  {\it a posteriori\/}, after the path integral has been evaluated, by evaluating the self-consistency conditions
\bea
G_c(\tau - \tau') &=& -\frac{1}{M M'} \sum_{\ell,\alpha} \left\langle c_{i\ell \alpha} (\tau) c_{i\ell \alpha}^\dagger (\tau') \right\rangle=  -\frac{1}{M M'} \sum_{\ell,\alpha} \left\langle b^\dagger_{\ell} (\tau) f_{ \alpha} (\tau) f^\dagger_{\alpha} (\tau') b_{\ell} (\tau') \right\rangle_\mathcal{Z} \nn \\
Q(\tau - \tau') &=& 
\frac{1}{M^2} \sum_{\alpha,\beta} \left\langle f_{i\alpha}^\dagger (\tau) f_{i\beta} (\tau) f_{i\beta}^\dagger (\tau') f_{i\alpha} (\tau') \right\rangle =
\frac{1}{M^2} \sum_{\alpha,\beta} \left\langle f_{\alpha}^\dagger (\tau) f_{\beta} (\tau) f_{\beta}^\dagger (\tau') f_{\alpha} (\tau') \right\rangle_\mathcal{Z} \,.
\label{saddle}
\eea

\subsection{Large $M$ limit}
\label{sec:largeM}

We have set things up to we can decouple the quartic terms in $\mathcal{S}$ by Hubbard-Stratonovich fields, perform the path integrals over $f_\alpha$ and $b_{\ell}$, and obtain a well-defined saddle point. See Refs.~\onlinecite{Fu18,Joshi2019}. The main difference from Ref.~\onlinecite{Fu18} is that both the boson and fermion Green's functions will have to be particle-hole asymmetric, as in Refs.~\onlinecite{SY92,GPS00,GPS01,SS15,Joshi2019}. The saddle-point equations for the slave particles $b_{\ell}$ and $f_\alpha$ are given as follows,
\begin{eqnarray}\label{Eq:EoM1}
  G_b(i\omega_n) &=& \frac{1}{i\omega_n+\mu_b-\Sigma_b(i\omega_n)} \\
  \Sigma_b(\tau) &=& -t^2G_f(\tau)G_f(-\tau)G_b(\tau) \label{Eq:EoM2}\\
  G_f(i\omega_n) &=& \frac{1}{i\omega_n+\mu_f-\Sigma_f(i\omega_n)} \label{Eq:EoM3}\\
  \Sigma_f(\tau) &=& -J^2 G_f(\tau)^2G_f(-\tau)+kt^2G_f(\tau)G_b(\tau)G_b(-\tau) \label{Eq:EoM4} 
\end{eqnarray}
Here $\mu_f$ and $\mu_b$ are chemical potentials, determined by $\epsilon_0$ and the saddle point value of $\lambda$, and chosen to satisfy
\beq
\left\langle f^\dagger f \right\rangle = \frac{1}{2}-k p \quad, \quad  \left\langle b^\dagger b \right\rangle = p \,. \label{lutt1}
\eeq
The electron Green's functions is a product  
 \begin{equation}
     G_c (\tau) = - G_f (\tau) G_b (-\tau)\,.
  \label{Eq:EoM5}
 \end{equation}

Eqs.\eqref{Eq:EoM1}-\eqref{Eq:EoM4} admit the following infra-red (IR) conformal saddle point solution at zero temperature \cite{Joshi2019}
\begin{equation}\label{eq:gft=0}
\begin{split}
   G_{a}^c(z) &= C_a \frac{e^{-i(\pi\Delta_a+\theta_a)\sgn\Im z}}{z^{1-2\Delta_a}}, \\
   \Sigma_{a}^c(z)&=-G_a^c(z)^{-1},\\
  G_{a}^c(\tau) &= -\frac{C_a\Gamma(2\Delta_a)}{\pi |\tau|^{2\Delta_a}}\sin(\theta_a+\pi\Delta_a\sgn\tau),\\
  \Sigma_{a}^c(\tau) &=-\frac{1}{C_a}\frac{\Gamma(2-2\Delta_a)}{\pi}\frac{1}{|\tau|^{2-2\Delta_a}}\sin(\theta_a+\pi\Delta_a\sgn\tau),\\
\end{split}
\end{equation}where $c$ in the superscript means conformal and subscript $a=b,f$ indexes boson and fermion respectively.

We will restrict our attention here to the the case where the boson and fermion scaling dimensions are $\Delta_f=\Delta_b=1/4$, which is the case proposed for the deconfined critical point in Ref.~\onlinecite{Joshi2019}.
The large $M$ equations also admit a solution in a critical phase \cite{Joshi2019} with $1/4 < \Delta_f < 1/2$, $\Delta_f + \Delta_b = 1/2$ which we will not explicitly consider here: we expect similar results to also apply to this critical phase. For the case $\Delta_f=\Delta_b = 1/4$,
the pre-factors satisfy
\begin{equation}\label{eq:CfCb}
  \begin{split}
     &t^2 C_f^2 C_b^2\cos(2\theta_f)  =\pi, \\
       & J^2 C_f^4 \cos(2\theta_f)-kt^2 C_f^2C_b^2\cos(2\theta_b)=\pi.
  \end{split}
\end{equation}
Also, from (\ref{lutt1}), $\theta_f,\theta_b$ satisfy the following Luttinger constraints \cite{GPS01,GKST2019}
\begin{eqnarray}
  \frac{\theta_f}{\pi}+\left(\frac{1}{2}-\Delta_f\right)\frac{\sin(2\theta_f)}{\sin(2\pi\Delta_f)} &=& kp, \label{eq:Ltf}\\
  \frac{\theta_b}{\pi}+\left(\frac{1}{2}-\Delta_b\right)\frac{\sin(2\theta_b)}{\sin(2\pi\Delta_b)} &=& \frac{1}{2}+p\,. \label{eq:LtB}
\end{eqnarray}
Along with the positivity constraints on the spectral weights, we then have the restrictions $-\pi\Delta_f<\theta_f<\pi\Delta_f$, $\pi\Delta_b<\theta_b<\pi/2$. 
We also remind here that $C_{b}$ and $C_{f}$ are defined to be real positive numbers, which requires 
\begin{equation}
    1+k \frac{\cos (2\theta_b)}{\cos (2\theta_f)}= \frac{J^2 C_f^4}{\pi} \cos (2\theta_f) > 0\,,
\end{equation}
where the first equality is due to \eqref{eq:CfCb}, and we have used $\Delta_f=1/4$ to constrain the range of the $\cos (2\theta_f)$ factor in the second step.\footnote{Remarkably, the same constraint will reappear later in section~\ref{sec:sta} for the stability of the conformal solution at $\Delta_f=\Delta_b=1/4$ from the resonance theory perspective.}

The above formalism applies also to the metal-insulator transition studied in Ref.~\onlinecite{Grisha2020} at half-filling, $p=0$. This corresponds to the case $\theta_f = 0$ and $\theta_b = \pi/2$.

\section{Formula for conductivity}
\label{sec:conductivity}

Our analysis of the large $z$ limit has so far only dealt with on-site Green's functions. To compute transport, we need the Green's function as a function of the lattice momentum, ${\bf k}$.
Using Eqns (7), (12), (13), (22) of Ref.~\onlinecite{Georges_RMP}, we can write down the momentum-dependent Green's function of the electrons
\beq
G_c ( {\bf k}, i\omega_n) = \frac{1}{ \displaystyle -\frac{\epsilon_{{\bf k}}}{\sqrt{M}} + \frac{t^2}{M} G_c (i\omega_n) + \frac{1}{G_c (i\omega_n)}} \label{Gk}
\eeq
where $\epsilon_{{\bf k}}$ is the dispersion on the lattice with hopping $t/\sqrt{z}$ (the local electron Green's function, $G_c (i \omega_n)$, is the momentum integral of $G_c ({\bf k}, i \omega_n)$).
Because $G_c \sim \mathcal{O}(M^0)$ in our large $M$ limit, the three terms in the denominator of (\ref{Gk}) all scale with different powers of $M$. To leading order in large $M$, we have the simple result
\beq
G_c ( {\bf k}, i\omega_n) \approx G_c (i\omega_n). \label{Gklocal}
\eeq
This momentum-independent form of the electron Green's function is quite distinct from the alternative large $M$ limit taken by Parcollet and Georges \cite{PG98}: we recall their results in \ref{app:PG}.

We can now write down the conductivity from the Kubo formula
\beq
\begin{split}
&i \omega \sigma(\omega) = \frac{2 M' e^2 t^2 a^{2-d}}{z}\int \frac{d \Omega_1}{2\pi} \frac{d \Omega_2}{2\pi}  A_c (\Omega_1) A_c (\Omega_2) \frac{ n_F (\Omega_2) - n_F(\Omega_1)}{ \Omega_1 - \Omega_2 - \omega - i \eta}\,,
\end{split}
\eeq
where $a$ is the lattice spacing, and $A_c (\Omega) = -2\mbox{Im} [ G_c (\Omega + i \eta)]$ is the spectral weight.
For the DC conductivity, we have
\beq
\sigma_{DC} = \frac{ M' e^2 t^2 a^{2-d}}{z} \int \frac{d \Omega}{2\pi}  A_c (\Omega)^2 \left( - \frac{\partial n_F}{\partial \Omega} \right)\,, \label{sigmaDC}
\eeq
where $d$ is the spatial dimension.
Note that $A_c (\Omega)$ is also implicitly a function of $T$.

We now evaluate (\ref{sigmaDC}) using the leading conformal solution presented in Section~\ref{sec:largeM}, we find at leading order the conductivity $\sigma_0$ is $T$-independent (see Eq.~\eqref{eq:sigma0}). To obtain a $T$-dependent resistivity, it's necessary to include corrections to the conformal solution (Sec.~\ref{sec:conformalcorrection}), which yields a correction
\begin{equation}\label{}
  \sigma_1=-8\sigma_0\alpha_G\frac{T}{J},
\end{equation} 
and therefore a linear in $T$ resistivity 
\begin{equation}\label{}
  \rho(T)=\rho(0)+\zeta T,
\end{equation} with slope $\zeta=8\rho(0)\alpha_G/J$.
Here $\alpha_G$ is the coefficient of correction to conformal Green's function (see Eq.~\eqref{eq:alphaG}). The exact value of $\alpha_G$ involves the ultra-violet (UV) behavior and can only be determined by numerics. However, using arguments of \cite{kitaev2017}, we will show that (Sec.~\ref{sec:eval_conductivity}) the slope can be related to the heat capacity $C=\gamma T$ by the relation \eqref{eq:zetagammarelation}
where the function
\begin{equation}\label{defgk}
  g(k,\theta_f,\theta_b)=-\frac{48}{\pi  \left[ 
  2(k\cos 2\theta_b -\cos 2\theta_f) +3\pi (k \cos^2 2\theta_b -\cos^2 2\theta_f)
  \right]
  }
\end{equation} 
can be determined entirely by $k$ and doping $p$ by the Luttinger constraints \eqref{eq:Ltf}, \eqref{eq:LtB}, and is therefore independent of UV details. To make a connection with the numerical study of the metal-insulator transition in \cite{Cha19}, we look at the half filling case $p=0$, $k=1/2$, $\theta_f=0$ and $\theta_b = \pi/2$ \cite{Grisha2020}, and we obtain $g={96}/({\pi  (6+3 \pi )})=1.981$.

\section{Perturbations around conformality}\label{sec:conformalcorrection}

This section turns to an examination of the corrections to the conformal ansatz solutions of Eqs. \eqref{Eq:EoM1}-\eqref{Eq:EoM4}. Such an analysis is aided by writing down an action whose saddle-point equations coincide with the Schwinger-Dyson 
Eqs. \eqref{Eq:EoM1}-\eqref{Eq:EoM4}. 
It is important to note that we will {\it not\/} be considering $1/M$ fluctuations around the saddle point of this action, just the corrections to the conformal ansatz {\it at\/} the saddle point. Fluctuations around the saddle point with the action below will not describe the $1/M$ corrections for the model considered in our paper. Instead, such fluctuations describe SYK models with random couplings which depend upon 4 indices (in contrast to the 2-index randomness considered here).
However, for the purposes of determining the deviation from conformality of the solutions of the saddle-point equations, the present action formulation turns out to be very useful. We will then be able to apply the soft-mode formalism developed for the SYK models \cite{kitaev2015talk,Maldacena2016,Bagrets2016,kitaev2017,GKST2019}.
\subsection{$G$-$\Sigma$ action}
We begin by writing down a $G$-$\Sigma$ action (per site, per spin) which reproduces the equations of motion in Eqs.\eqref{Eq:EoM1}-\eqref{Eq:EoM4}\footnote{In contrast to the SYK model, where the $G$-$\Sigma$ action is an exact rewriting of the model. Here in this paper we only treat the $G$-$\Sigma$ action as a handy tool to analyze the Schwinger-Dyson equation. Alternatively, one can achieve the same conclusions by investigating the Schwinger-Dyson equation only. For example, the kernel $W_{G}$ and $W_{\Sigma}$ (see \eqref{Eq:W}) that will be the center of our analysis are well-defined in the Schwinger-Dyson setting without referring to the action.}
\begin{equation}\label{Eq:S}
\begin{split}
  S&=k\log\det\left((\partial_\tau-\mu_b)\delta(\tau_1-\tau_2)+\Sigma_b(\tau_1,\tau_2)\right)-\log\det\left((\partial_\tau-\mu_f)\delta(\tau_1-\tau_2)+\Sigma_f(\tau_1,\tau_2)\right)\\
  &+k \Tr(\Sigma_b\cdot G_b)-\Tr(\Sigma_f\cdot G_f)+ \frac{k t^2}{2}\Tr\left((G_f G_b)\cdot (G_f G_b)\right)-\frac{J^2}{4}\Tr\left(G_f^2\cdot G_f^2\right).
\end{split}
\end{equation}
Here we use the notations in Ref.~\onlinecite{GKST2019}:
\begin{equation}
  \Tr(f\cdot g)\equiv\int\rd \tau_1\rd \tau_2 f(\tau_2,\tau_1)g(\tau_1,\tau_2), \quad
  f^T A g\equiv\int \rd^4 \tau f(\tau_2,\tau_1)A(\tau_1,\tau_2;\tau_3,\tau_4)g(\tau_3,\tau_4).
\end{equation}
In \eqref{Eq:S} the $\delta$-function terms are UV sources because their spectra span all frequencies. To make the symmetries of \eqref{Eq:S} manifest and the perturbation theory around the conformal solution transparent, we shift the self energies by $\Sigma_a\to\Sigma_a-\sigma_a$, $\sigma_a(\tau_1,\tau_2)=(\partial_\tau-\mu_a)\delta(\tau_1-\tau_2)$ :
\begin{equation}\label{Eq:S2}
\begin{split}
  S&=k\log\det\left(\Sigma_b(\tau_1,\tau_2)\right)-\log\det\left(\Sigma_f(\tau_1,\tau_2)\right)+k \Tr(\Sigma_b\cdot G_b)-\Tr(\Sigma_f\cdot G_f)\\
  &+ \frac{k t^2}{2}\Tr\left((G_f G_b)\cdot (G_f G_b)\right)-\frac{J^2}{4}\Tr\left(G_f^2\cdot G_f^2\right)-k\Tr(\sigma_b\cdot G_b)+\Tr(\sigma_f\cdot G_f).
\end{split}
\end{equation}
In absence of the UV sources $\sigma_b,\sigma_f$, the action \eqref{Eq:S2} has the following U(1) and reparameterization symmetries
\begin{equation}\label{eq:symmu1}
\begin{split}
  G_a(\tau_1,\tau_2)&\to G_a(\tau_1,\tau_2)e^{i(\lambda(\tau_1)-\lambda(\tau_2))},\\
  \Sigma_a(\tau_1,\tau_2)&\to \Sigma_a(\tau_1,\tau_2)e^{i(\lambda(\tau_1)-\lambda(\tau_2))},
\end{split}
\end{equation}
\begin{equation}\label{eq:symmrep}
\begin{split}
   G_a(\tau_1,\tau_2) & \to G_a(f(\tau_1),f(\tau_2))f'(\tau_1)^{\Delta_a}f'(\tau_2)^{\Delta_b}, \\
   \Sigma_a(\tau_1,\tau_2) & \to \Sigma_a(f(\tau_1),f(\tau_2))f'(\tau_1)^{1-\Delta_a}f'(\tau_2)^{1-\Delta_b},
\end{split}
\end{equation} where $a=b,~f$ and $\Delta_b=\Delta_f=1/4$. The invariance of the $\det$ terms demands a regularization by free particle Green's function, see \cite{GKST2019,kitaev2017}.
The insertion of UV sources explicitly breaks the above symmetries. As we will see later, the (quasi-) soft modes arising from the above symmetries are dual to the leading corrections to conformality in the infra-red (IR).

\subsection{Linear response to UV sources}

The correction to conformality in presence of the UV sources $\sigma_b,~\sigma_f$ can now be formulated as a linear response problem by assuming $\sigma_b,\sigma_f$ are small in the IR region $|\tau|\gg J^{-1}$ (we treat $J$ and $t$ to be of the same order and choose to use $J$ to represent that scale in the paper). The rationale behind this assumption is the RG flow picture discussed in \cite{kitaev2017,GKST2019}: for a generic UV source $\sigma$, its response will be concentrated at short timescales $|\tau|\lesssim J^{-1}$, but there are some resonant perturbations which is able to produce influence in the IR of the following form ($\beta \equiv 1/T$)
\begin{equation}\label{eq:deltaGschematic}
  \delta G_a(\tau)\propto |J \tau|^{1-h} G_{a}^c(\tau),\quad J^{-1}\ll |\tau|\ll \beta,
\end{equation} with certain exponent $h\geq1$. Dimensional analysis shows that such a term will be of order $(\beta J)^{1-h}$. Therefore for $h>1$ such terms are small at low temperatures and can be treated within linear response. The $h=1$ correction doesn't fit in this argument, but we will show that they are equivalent to changing $\theta_f,\theta_b$ in \eqref{eq:gft=0} and can be absorbed into these parameters. We will first work at zero temperature limit $\beta=\infty$ and later relax the $|\tau|\ll \beta$ condition in \eqref{eq:deltaGschematic} and derive a result applicable at finite temperature.

Following the spirit of linear response, we expand the action \eqref{Eq:S2} around the conformal saddle point $G=G^{c}+\delta G,~\Sigma=\Sigma^c+\delta\Sigma$ to quadratic order in $\delta G,~\delta\Sigma,~\sigma$:
\begin{equation}\label{Eq:deltaS}
 \delta S=\frac{1}{2}\begin{pmatrix}
                 \delta\Sigma^T & \delta G^T
               \end{pmatrix}
                \Lambda
               \begin{pmatrix}
                 W_\Sigma & -1 \\
                 -1 & W_G
               \end{pmatrix}
               \begin{pmatrix}
                 \delta\Sigma \\
                 \delta G
               \end{pmatrix}+\Tr(\delta G\cdot\Lambda \sigma).
\end{equation}
Here $\delta \Sigma$, $\delta G$ and $\sigma$ should be understood as two-component vectors in $(b,f)$ space, and $\Lambda=\text{diag}(-k,1)$ is a matrix in $(b,f)$ space.

The kernels $W_\Sigma$ and $W_G$ are defined as 
\begin{equation}
  W_\Sigma = \frac{\delta G^c[\Sigma^c]}{\delta \Sigma^c}, \quad 
  W_G = \frac{\delta \Sigma^c[G^c]}{\delta G^c},\label{Eq:W}
\end{equation} where $G^c[\Sigma^c]$ means we plug in the Schwinger-Dyson  equations for $G^c$ (eqs.\eqref{Eq:EoM1}, \eqref{Eq:EoM3}) with UV terms dropped) and evaluate the functional derivative at saddle point; a similar meaning applies to $\Sigma^c[G^c]$ (eqs. \eqref{Eq:EoM2}, \eqref{Eq:EoM4}). 
Restoring the subscripts $(b,f)$ and time indices, the above equations become
\begin{equation}
    \begin{aligned}
    W_{\Sigma} (\tau_1,\tau_2;\tau_3,\tau_4)_{a\tilde{a}} &= \frac{\delta G^c_{a}(\tau_1,\tau_2)}{\delta \Sigma^c_{\tilde{a}}(\tau_3,\tau_4)} \\
    W_{G} (\tau_1,\tau_2;\tau_3,\tau_4)_{a\tilde{a}} &= \frac{\delta \Sigma^c_{a}(\tau_1,\tau_2)}{\delta G^c_{\tilde{a}}(\tau_3,\tau_4)}
    \end{aligned}
    \label{Eq:Windex}
\end{equation}
Minimizing \eqref{Eq:deltaS} with respect to $\delta G,~\delta\Sigma$, we obtain the following response to UV sources:
\begin{equation}
    \delta G = (1-\underbrace{W_\Sigma W_G}_{K_G})^{-1}W_\Sigma \sigma \,, \quad
  \delta \Sigma =  (1-\underbrace{W_G W_\Sigma}_{K_\Sigma})^{-1} \sigma\,.
\label{eq:deltaGsigma}
\end{equation}
Such a form of response indicates that for a source $\sigma$ supported in the UV, only if $(1-K_G)^{-1}$ and $(1-K_\Sigma)^{-1}$ are singular can there be IR responses (resonance). Resonance happens both in $\delta G$ and $\delta \Sigma$ because an eigenvector $v$ of $K_\Sigma$ with unit eigenvalue is in one-to-one correspondence to an eigenvector $W_\Sigma v$ of $K_G$ with the same eigenvalue. Therefore our task is to find resonant eigenvectors of $K_G,~K_\Sigma$.

\subsection{Explicit computation of $W_\Sigma$, $W_G$ at zero temperature}


Using Eq.\eqref{Eq:Windex} and the conformal saddle point equations, we obtain
\begin{equation}\label{}
  W_\Sigma(1,2;3,4)=\begin{pmatrix}
                      G_b^{13}G_b^{42} & 0 \\
                      0 & G_f^{13}G_f^{42}
                    \end{pmatrix},
\end{equation} where we have abbreviated the time argument $(\tau_i,\tau_j)$ by the superscript $ij$, and the two rows denote $b,f$ respectively. The Green's functions take the conformal saddle point value \eqref{eq:gft=0}.
Diagrammatically, $W_{\Sigma}$ has the following representation
\begin{equation}\label{eq:Wsigmagraph}
  W_\Sigma(1,2;3,4)=\begin{pmatrix}
                      \begin{tikzpicture}[baseline={([yshift=-4pt]current bounding box.center)}]
                     \draw[thick, dashed, mid arrow] (40pt,12pt)--(0pt,12pt);
                     \draw[thick, dashed, mid arrow] (0pt,-12pt)--(40pt,-12pt);
                     \node at (-5pt,12pt) {\scriptsize $1$};
                     \node at (-5pt,-12pt) {\scriptsize $2$};
                     \node at (48pt,12pt) {\scriptsize $3$};
                     \node at (48pt,-12pt) {\scriptsize $4$};
                     \end{tikzpicture}
                        & 0 \\
                      0 & 
                      \begin{tikzpicture}[baseline={([yshift=-4pt]current bounding box.center)}]
                     \draw[thick, mid arrow] (40pt,12pt)--(0pt,12pt);
                     \draw[thick, mid arrow] (0pt,-12pt)--(40pt,-12pt);
                     \node at (-5pt,12pt) {\scriptsize $1$};
                     \node at (-5pt,-12pt) {\scriptsize $2$};
                     \node at (48pt,12pt) {\scriptsize $3$};
                     \node at (48pt,-12pt) {\scriptsize $4$};
                     \end{tikzpicture}
                    \end{pmatrix}.
\end{equation}
An arrowed dashed line denotes $G_b$ , and solid line denotes $G_f$.
The result for $W_G$ is
\begin{equation}\label{}
 W_G(1,2;3,4)=
               \begin{pmatrix}
                 -t^2 G_f^{12}G_f^{21} \delta^{24;13}& -t^2 G_f^{43} G_b^{12}(\delta^{24;13}+\delta^{23;14})   \\
                 kt^2 G_b^{43}G_f^{12}(\delta^{24;13}+\delta^{23;14}) 
                 & 
                 kt^2 G_b^{12} G_b^{21}\delta^{24;13}-J^2\left(2 G_f^{12}G_f^{21}\delta^{24;13}+(G_f^{12})^2\delta^{23;14}\right)
               \end{pmatrix}
\end{equation} 
where notation $\delta^{ij;kl}=\delta(\tau_i-\tau_j)\delta(\tau_k-\tau_l)$. Diagrammatically, $W_G$ has the following representation
\begin{equation}\label{}
  W_G(1,2;3,4)=\begin{pmatrix}
        -t^2
        \begin{tikzpicture}[baseline={([yshift=-4pt]current bounding box.center)}]
        \draw[thick, densely dotted] (10pt,15pt)--(0pt,15pt);
        \draw[thick, densely dotted] (0pt,-15pt)--(10pt,-15pt);
        \draw[thick, mid arrow] (0pt,15pt)..controls (5pt,7pt) and (5pt,-7pt)..(0pt,-15pt);
        \draw[thick, mid arrow] (0pt,-15pt)..controls (-5pt,-7pt) and (-5pt,7pt)..(0pt,15pt);
        \node at (-5pt,15pt) {\scriptsize $1$};
        \node at (-5pt,-15pt) {\scriptsize $2$};
        \node at (18pt,15pt) {\scriptsize $3$};
        \node at (18pt,-15pt) {\scriptsize $4$};
        \end{tikzpicture}
& -t^2\left(
         \begin{tikzpicture}[baseline={([yshift=-4pt]current bounding box.center)}]
        \draw[thick, densely dotted] (10pt,15pt)--(0pt,15pt);
        \draw[thick, densely dotted] (0pt,-15pt)--(10pt,-15pt);
        \draw[thick,  mid arrow] (0pt,15pt)..controls (5pt,7pt) and (5pt,-7pt)..(0pt,-15pt);
        \draw[thick, dashed, mid arrow] (0pt,-15pt)..controls (-5pt,-7pt) and (-5pt,7pt)..(0pt,15pt);
        \node at (-5pt,15pt) {\scriptsize $1$};
        \node at (-5pt,-15pt) {\scriptsize $2$};
        \node at (18pt,15pt) {\scriptsize $3$};
        \node at (18pt,-15pt) {\scriptsize $4$};
        \end{tikzpicture}
        +
        \begin{tikzpicture}[baseline={([yshift=-4pt]current bounding box.center)}]
        \draw[thick, densely dotted] (0pt,15pt)--(20pt,-15pt);
        \draw[thick, densely dotted] (20pt,15pt)--(0pt,-15pt);
        \draw[thick,  mid arrow] (0pt,-15pt)..controls (5pt,-7pt) and (5pt,7pt)..(0pt,15pt);
        \draw[thick, dashed,mid arrow] (0pt,-15pt)..controls (-5pt,-7pt) and (-5pt,7pt)..(0pt,15pt);
        \node at (-5pt,15pt) {\scriptsize $1$};
        \node at (-5pt,-15pt) {\scriptsize $2$};
        \node at (28pt,15pt) {\scriptsize $3$};
        \node at (28pt,-15pt) {\scriptsize $4$};
        \end{tikzpicture}\right)  \vspace{1em}\\
        kt^2\left(
         \begin{tikzpicture}[baseline={([yshift=-4pt]current bounding box.center)}]
        \draw[thick, densely dotted] (10pt,15pt)--(0pt,15pt);
        \draw[thick, densely dotted] (0pt,-15pt)--(10pt,-15pt);
        \draw[thick, dashed, mid arrow] (0pt,15pt)..controls (5pt,7pt) and (5pt,-7pt)..(0pt,-15pt);
        \draw[thick, mid arrow] (0pt,-15pt)..controls (-5pt,-7pt) and (-5pt,7pt)..(0pt,15pt);
        \node at (-5pt,15pt) {\scriptsize $1$};
        \node at (-5pt,-15pt) {\scriptsize $2$};
        \node at (18pt,15pt) {\scriptsize $3$};
        \node at (18pt,-15pt) {\scriptsize $4$};
        \end{tikzpicture}
        +
        \begin{tikzpicture}[baseline={([yshift=-4pt]current bounding box.center)}]
        \draw[thick, densely dotted] (0pt,15pt)--(20pt,-15pt);
        \draw[thick, densely dotted] (20pt,15pt)--(0pt,-15pt);
        \draw[thick, dashed, mid arrow] (0pt,-15pt)..controls (5pt,-7pt) and (5pt,7pt)..(0pt,15pt);
        \draw[thick, mid arrow] (0pt,-15pt)..controls (-5pt,-7pt) and (-5pt,7pt)..(0pt,15pt);
        \node at (-5pt,15pt) {\scriptsize $1$};
        \node at (-5pt,-15pt) {\scriptsize $2$};
        \node at (28pt,15pt) {\scriptsize $3$};
        \node at (28pt,-15pt) {\scriptsize $4$};
        \end{tikzpicture}\right) 
        \hspace{10pt}
    & kt^2 
        \begin{tikzpicture}[baseline={([yshift=-4pt]current bounding box.center)}]
        \draw[thick, densely dotted] (10pt,15pt)--(0pt,15pt);
        \draw[thick, densely dotted] (0pt,-15pt)--(10pt,-15pt);
        \draw[thick, dashed,mid arrow] (0pt,15pt)..controls (5pt,7pt) and (5pt,-7pt)..(0pt,-15pt);
        \draw[thick, dashed,mid arrow] (0pt,-15pt)..controls (-5pt,-7pt) and (-5pt,7pt)..(0pt,15pt);
        \node at (-5pt,15pt) {\scriptsize $1$};
        \node at (-5pt,-15pt) {\scriptsize $2$};
        \node at (18pt,15pt) {\scriptsize $3$};
        \node at (18pt,-15pt) {\scriptsize $4$};
        \end{tikzpicture}
        -J^2
        \left(2
         \begin{tikzpicture}[baseline={([yshift=-4pt]current bounding box.center)}]
        \draw[thick, densely dotted] (10pt,15pt)--(0pt,15pt);
        \draw[thick, densely dotted] (0pt,-15pt)--(10pt,-15pt);
        \draw[thick,  mid arrow] (0pt,15pt)..controls (5pt,7pt) and (5pt,-7pt)..(0pt,-15pt);
        \draw[thick, mid arrow] (0pt,-15pt)..controls (-5pt,-7pt) and (-5pt,7pt)..(0pt,15pt);
        \node at (-5pt,15pt) {\scriptsize $1$};
        \node at (-5pt,-15pt) {\scriptsize $2$};
        \node at (18pt,15pt) {\scriptsize $3$};
        \node at (18pt,-15pt) {\scriptsize $4$};
        \end{tikzpicture}
        +
        \begin{tikzpicture}[baseline={([yshift=-4pt]current bounding box.center)}]
        \draw[thick, densely dotted] (0pt,15pt)--(20pt,-15pt);
        \draw[thick, densely dotted] (20pt,15pt)--(0pt,-15pt);
        \draw[thick,  mid arrow] (0pt,-15pt)..controls (5pt,-7pt) and (5pt,7pt)..(0pt,15pt);
        \draw[thick, mid arrow] (0pt,-15pt)..controls (-5pt,-7pt) and (-5pt,7pt)..(0pt,15pt);
        \node at (-5pt,15pt) {\scriptsize $1$};
        \node at (-5pt,-15pt) {\scriptsize $2$};
        \node at (28pt,15pt) {\scriptsize $3$};
        \node at (28pt,-15pt) {\scriptsize $4$};
        \end{tikzpicture}\right)  \vspace{1em}
      \end{pmatrix},\label{eq:WGgraphic}
\end{equation}where a dotted line without arrow represents a $\delta$-function. 
Note $W_{\Sigma}$ alone is not symmetric due to the additional $k$ factor. However, the product
\begin{equation}
  \Lambda  \begin{pmatrix}
    W_{\Sigma} & -1 \\
    -1 & W_{G}
    \end{pmatrix} 
\end{equation}
is symmetric as required by the quadratic expansion. 

At zero temperature, all eigenfunctions of $K_G$ and $K_\Sigma$ are SL(2,$R$) covariant and therefore are powerlaws. This motivates the following representation of $\sigma$, $\delta\Sigma$ and $\delta G$:
 \begin{equation}\label{eq:sigmaansatz}
   \sigma(\tau)=\begin{pmatrix}
                   \delta \sigma_{b+} \Sigma_{b}^c(\tau) \\
                   \delta \sigma_{b-} \Sigma_{b}^c(\tau)\\
                   \delta \sigma_{f+} \Sigma_{f}^c(\tau)\\
                   \delta \sigma_{f-} \Sigma_{f}^c(\tau),
                 \end{pmatrix}|\tau|^{1-h}      
 \end{equation}
\begin{equation}\label{eq:ansatz}
  \delta G(\tau)=\begin{pmatrix}
                   \delta G_{b+} G_{b}^c(\tau) \\
                   \delta G_{b-} G_{b}^c(\tau)\\
                   \delta G_{f+} G_{f}^c(\tau)\\
                   \delta G_{f-} G_{f}^c(\tau)
                 \end{pmatrix}|\tau|^{1-h},
                 \qquad
    \delta \Sigma(\tau)=\begin{pmatrix}
                   \delta \Sigma_{b+} \Sigma_{b}^c(\tau) \\
                   \delta \Sigma_{b-} \Sigma_{b}^c(\tau)\\
                   \delta \Sigma_{f+} \Sigma_{f}^c(\tau)\\
                   \delta \Sigma_{f-} \Sigma_{f}^c(\tau)
                 \end{pmatrix}|\tau|^{1-h}             .
\end{equation}
Here $\pm$ means positive/negative $\tau$ branches, i.e. $\sigma(\tau)$ is a powerlaw with different prefactors at $\tau>0$ and $\tau<0$.

The above conformal ansatz will be preserved by $W_\Sigma$ and $W_G$, and our next goal is to work out the matrix representations $W_\Sigma(h)$, $W_G(h)$ acting on the coefficients $\delta G_{a\pm},\delta \Sigma_{a\pm}$.
The computation of $W_\Sigma(h)$ is identical to \cite{GKST2019}, yielding
\begin{equation}
\begin{split}
    W_\Sigma(h)&=w(\Delta_b,\theta_b;h)\oplus w(\Delta_f,\theta_f;h),\\
    w(\Delta,\theta;h)&=\frac{\Gamma(2\Delta-1+h)\Gamma(2\Delta-h)}{\Gamma(2\Delta)\Gamma(2\Delta-1)\sin(2\pi\Delta)}
    \begin{pmatrix}
                                                                                                               \sin(\pi h+2\theta) & -\sin(2\pi\Delta)+\sin(2\theta) \\
                                                                                                               -\sin(2\pi\Delta)-\sin(2\theta) & \sin(\pi h-2\theta)
                                                                                                             \end{pmatrix},
\end{split}
\end{equation}
$W_G$ is computed by inserting the ansatz \eqref{eq:ansatz} into \eqref{eq:WGgraphic} and using the Schwinger-Dyson equation, we obtain
\begin{equation}\label{}
  W_G(h)=\begin{pmatrix}
           1  & 0 & 1 & 1 \\
           0 & 1 & 1 & 1 \\
           \frac{kt^2 \rho \rho^T}{-J^2+k t^2 \rho \rho^T} & \frac{kt^2 \rho \rho^T}{-J^2+k t^2 \rho \rho^T} & \frac{-2J^2+kt^2 \rho \rho^T}{-J^2+k t^2 \rho \rho^T} & \frac{-J^2}{-J^2+k t^2 \rho\rho^T} \\
           \frac{kt^2 \rho \rho^T}{-J^2+k t^2 \rho \rho^T} & \frac{kt^2 \rho \rho^T}{-J^2+k t^2 \rho \rho^T} & \frac{-J^2}{-J^2+k t^2 \rho\rho^T} & \frac{-2J^2+kt^2 \rho \rho^T}{-J^2+k t^2 \rho \rho^T}
         \end{pmatrix}.
\end{equation}
Here $\rho,\rho^T$ are defined as
\begin{equation}\label{}
  G_{b}^c(\tau)=\rho(\tau)G_{f}^c(\tau),\qquad \rho^T(\tau)=\rho(-\tau).
\end{equation}
$\rho(\tau)$ is a piecewise constant function. Using \eqref{eq:CfCb}, it can be shown that
\begin{equation}\label{}
  \rho\rho^T=\frac{J^2}{t^2}\frac{\cos2\theta_b}{\cos2\theta_f+k\cos2\theta_b}.
\end{equation} 
Therefore
\begin{equation}\label{}
  W_G(h)=\left(
\begin{array}{cccc}
 1 & 0 & 1 & 1 \\
 0 & 1 & 1 & 1 \\
 -k\frac{\cos 2\theta_b}{\cos 2 \theta_f} & -k\frac{\cos 2\theta_b}{\cos 2 \theta_f}  & k \frac{\cos 2\theta_b}{\cos 2 \theta_f}+2 & k \frac{\cos 2\theta_b}{\cos 2 \theta_f}+1 \\
 -k\frac{\cos 2\theta_b}{\cos 2 \theta_f} & -k\frac{\cos 2\theta_b}{\cos 2 \theta_f} & k \frac{\cos 2\theta_b}{\cos 2 \theta_f}+1 & k \frac{\cos 2\theta_b}{\cos 2 \theta_f}+2 \\
\end{array}
\right),
\end{equation} which is independent of $h$.

From this we can compute $K_G(h)=W_\Sigma(h)W_G(h)$ and $K_\Sigma(h)=W_G(h)W_\Sigma(h)$. By construction $K_G(h)$ and $K_\Sigma(h)$ have the same nonzero spectrum. There is a further duality between $h$ and $1-h$ given by the identity 
\begin{equation}
  \Lambda  K_G(h)= P^{-1} K_\Sigma(1-h)^T \Lambda P,  \quad \text{where}~~
  P = \sigma_x \otimes {\rm diag}(1,\cos2\theta_f \sec2\theta_b). 
\end{equation} This is related to $(\Lambda K_G)_{a\tilde{a}}(1,2;3,4)=( K_\Sigma \Lambda )_{\tilde{a}a}(4,3;2,1)$, which comes from an 180-degree rotation of diagrammatic representations \eqref{eq:Wsigmagraph},\eqref{eq:WGgraphic}. We have generalized results of \cite{GKST2019} that $K_G(h)$, $K_\Sigma(h)$, $K_G(1-h)$ and $K_\Sigma(1-h)$ share the same nonzero spectrum.

\subsection{Resonances at zero temperature}\label{sec:Resonance}

Now we consider the resonances of $K_G(h)=W_\Sigma(h)W_G(h)$.
 Direct computation confirms that $K_G(h)$ has eigenvalue $1$ at $h=-1,0,1,2$, for any $k,\theta_f,\theta_b$, which is a manifestation of U(1) \eqref{eq:symmu1} and reparameterization symmetry \eqref{eq:symmrep}. There is also higher $h$ resonance but the value of $h$ is transcendental and it depends on $k,\theta_f,\theta_b$.

 In the following we will focus on the $h=2$ and $h=1$ resonances. We will derive the exact form of Green's function correction at $h=2$ using the RG scheme of \cite{kitaev2017}, and show that its coefficient $\alpha_G$ is related to the coefficient $\alpha_S$ of Schwarzian action by a linear relation. The proportionality of $\alpha_G$ to $\alpha_S$ is because the $h=2$ source couples to the IR reparameterization modes (which has scaling dimension -1). 
 
 We also argue that the $h=2$ resonance is the leading correction from UV sources. First, we will examine the $h=1$ resonance, and show that the correction can be absorbed into redefinitions of $\theta_f,\theta_b$. Second, we should argue that there is no other resonance between $0<h<2$. However, this is no guaranteed for arbitrary $(k,\theta_f,\theta_b)$ or $(k,p)$ (two sets of parameters are related by Luttinger constraints). Instead, we shall argue that the absence of resonance in $0<h<2$ other than $h=1$ is implied by stability of the conformal solution \eqref{eq:gft=0}, and work out the parameter range of $(k,p)$ allowed by stability.

\subsubsection{$h=2$ sector}

  In the $h=2$ sector, the kernel $K_G(h)$ has a pair of left and right eigenvectors
\begin{equation}\label{}
\begin{split}
  &v_2^L=\frac{1}{4(-k\cos2\theta_b+\cos2\theta_f)}\begin{pmatrix}
                                                   -k\cos2\theta_b & -k\cos2\theta_b & \cos2\theta_f & \cos2\theta_f
                                                 \end{pmatrix},\\
  &v_2^R=\begin{pmatrix}
        2-3\sin2\theta_b \\
        2+3\sin2\theta_b \\
        2-3\sin2\theta_f \\
        2+3\sin2\theta_f
      \end{pmatrix},\quad v_2^L v_2^R=1.
\end{split}
\end{equation}
Following \cite{kitaev2017}, we consider a regulated perturbation of the form
\begin{equation}\label{eq:sigmasource}
  \sigma(\tau)=\begin{pmatrix}
                   \sigma_{b+} \Sigma_{b}^c(\tau) \\
                   \sigma_{b-} \Sigma_{b}^c(\tau)\\
                    \sigma_{f+} \Sigma_{f}^c(\tau)\\
                   \sigma_{f-} \Sigma_{f}^c(\tau)
                 \end{pmatrix}|\tau|^{1-h} u(\ln|\tau|)\equiv\Sigma^c(\tau)\vec{\sigma}|\tau|^{1-h}u(\ln|\tau|),
\end{equation}where to regulate the divergence in $(1-K_G(h))^{-1}$ we have introduced a window function $u(\xi)$ of scale parameter $\xi=\ln|\tau|$. The second term is a short-hand of the first term. The window function is normalized such that
\begin{equation}\label{}
  \int_{-\infty}^{\infty} u(\xi) \rd \xi =1.
\end{equation}

Using \eqref{eq:deltaGsigma} and the same arguments as \cite{kitaev2017} , we can compute the responses of different Fourier components of $u(\xi)$ and recombine them together. Under the assumption that $u(\xi)$ is sufficiently wide, the response of Green's function at resonance $h=h_*$ is
\begin{equation}\label{eq:deltaGtw}
  \delta G(\tau)=\frac{1}{K_G'(h_*)} W_\Sigma(h_*)\vec{\sigma}G^c(\tau)|\tau|^{1-h_*}w(\ln|\tau|),
\end{equation}where
\begin{equation}
    w(\xi)=\int_{-\infty}^{\xi} u(\zeta) \rd \zeta \,.
\end{equation}
Here the $(K_G'(h_*))^{-1}$ term should be interpreted as
\begin{equation}\label{eq:kgpinverse}
  \frac{1}{K_G'(h_*)}=\frac{1}{k_G'(h_*)}v^R_{h_*}v^L_{h_*},
\end{equation} where $k_G(h_*)=1$ is the interested eigenvalue branch of $K_G(h_*)$ and $k_G'(h_*)$ is its derivative. 

Now set $h_*=2$, and we obtain
\begin{equation}\label{eq:kgp2}
  k_G'(2)=v^L_{2}K_G'(2)v^R_{2}=\frac{(1-k)\pi  k \cos ^22 \theta _b}{\cos 2 \theta _f-k \cos 2 \theta _b}-\pi\left(\cos 2\theta_f+k\cos 2\theta_b\right)-\frac{2}{3},
\end{equation}
and the response is
\begin{equation}\label{eq:alphaG}
\begin{split}
   &\delta G=-\frac{\alpha_G}{J}  \begin{pmatrix}
        2-3\sin2\theta_b\\
        2+3\sin2\theta_b \\
        2-3\sin2\theta_f \\
        2+3\sin2\theta_f
      \end{pmatrix}G^c(\tau)|\tau|^{-1}w(\ln|\tau|), \\
   &\frac{\alpha_G}{J}  = -\frac{v_2^L W_\Sigma(2)\vec{\sigma}}{ k_G'(2)}=-\frac{\cos 2 \theta _f \left(\sigma _{f-}+\sigma _{f+}\right)-k \cos2 \theta _b \left(\sigma _{b-}+\sigma _{b+}\right)}{12(\cos2\theta_f-k\cos2\theta_b)k_G'(2)}.
\end{split}
\end{equation}
Notice that for sufficiently large $|\tau|$, $w(\ln|\tau|)\approx 1$, and we recover the powerlaw form.

We remark that for the actual UV source $\sigma(\tau)=(\partial_\tau-\mu)\delta(\tau)$ we do not know how to decompose it into the form of \eqref{eq:sigmasource}, so we are not able to calculate the numerical value of $\alpha_G$ analytically, and its value can only be determined through fitting the numerical solution of the full Schwinger-Dyson equation.

Next we demonstrate that $\alpha_G$ is related to the coefficient of the Schwarzian action.
The $h=2$ source $\sigma(\tau)$ given in \eqref{eq:sigmasource} couples to the reparameterization modes in the IR:
\begin{equation}\label{eq:GIRSch}
 G_{\rm IR}(\tau_1,\tau_2)=G^c(f(\tau_1),f(\tau_2))f'(\tau_1)^{1/4}f'(\tau_2)^{1/4}\approx G^c(\tau_1,\tau_2)(1+\frac{1}{24}{\rm Sch}(e^{if(\tau_+)},\tau_+)\tau_{12}^2+\dots),
\end{equation} where in the second equality we expanded in terms of small $\tau_{12}=\tau_1-\tau_2$, and $\tau_+=(\tau_1+\tau_2)/2$.

If there were no UV sources, such a reparameterization costs no action, but turning on UV sources introduces an action cost
\begin{equation}\label{}
  \delta S=\Tr(\sigma\cdot\Lambda G),
\end{equation} and we can evaluate it on \eqref{eq:GIRSch} to get the action of reprameterization. The zeroth order term in $\tau_{12}$ introduces a shift in ground state energy, and the first order term yields the Schwarzian action
\begin{equation}\label{}
  \delta S=-\frac{\alpha_S}{J}\int\rd\tau_+{\rm Sch}(e^{if(\tau_+)},\tau_+),
\end{equation}
and the coefficient is
\begin{equation}\label{eq:alphaS}
\begin{split}
  \frac{\alpha_S}{J}&=\frac{1}{96\pi}\left[(\sigma_{f+}+\sigma_{f-})\cos2\theta_f-(\sigma_{b+}+\sigma_{b-})k\cos2\theta_b\right]\int_{0}^{\infty}\rd\tau_{12}|\tau_{12}|^{-1}u(\ln |\tau|)\\
                    &=\frac{1}{96\pi}\left[(\sigma_{f+}+\sigma_{f-})\cos2\theta_f-(\sigma_{b+}+\sigma_{b-})k\cos2\theta_b\right].
\end{split}
\end{equation}
Therefore we have the large $M$ relation
\begin{equation}\label{eq:aGaS}
  \frac{\alpha_G}{\alpha_S}=\frac{8\pi}{-k_G'(2)(\cos2\theta_f-k\cos2\theta_b)},
\end{equation}where $k_G'(2)$ is given in \eqref{eq:kgp2}. 

At $k=0$, our system reduces to a pure complex $\rm{SYK}_4$ model, and the above result agrees with the Majorana $\rm{SYK}_4$ result \cite{Maldacena2016,kitaev2017} up to a factor of 4, which can be explained by (i) our definition of $\alpha_G$ is smaller by a factor of 1/2; (ii) a complex SYK model has twice the degrees of freedom of the Majorana version, so our $\alpha_S$ should scale larger by a factor 2.

\begin{figure}[htb!]
  \centering
  \includegraphics[width=0.5\textwidth]{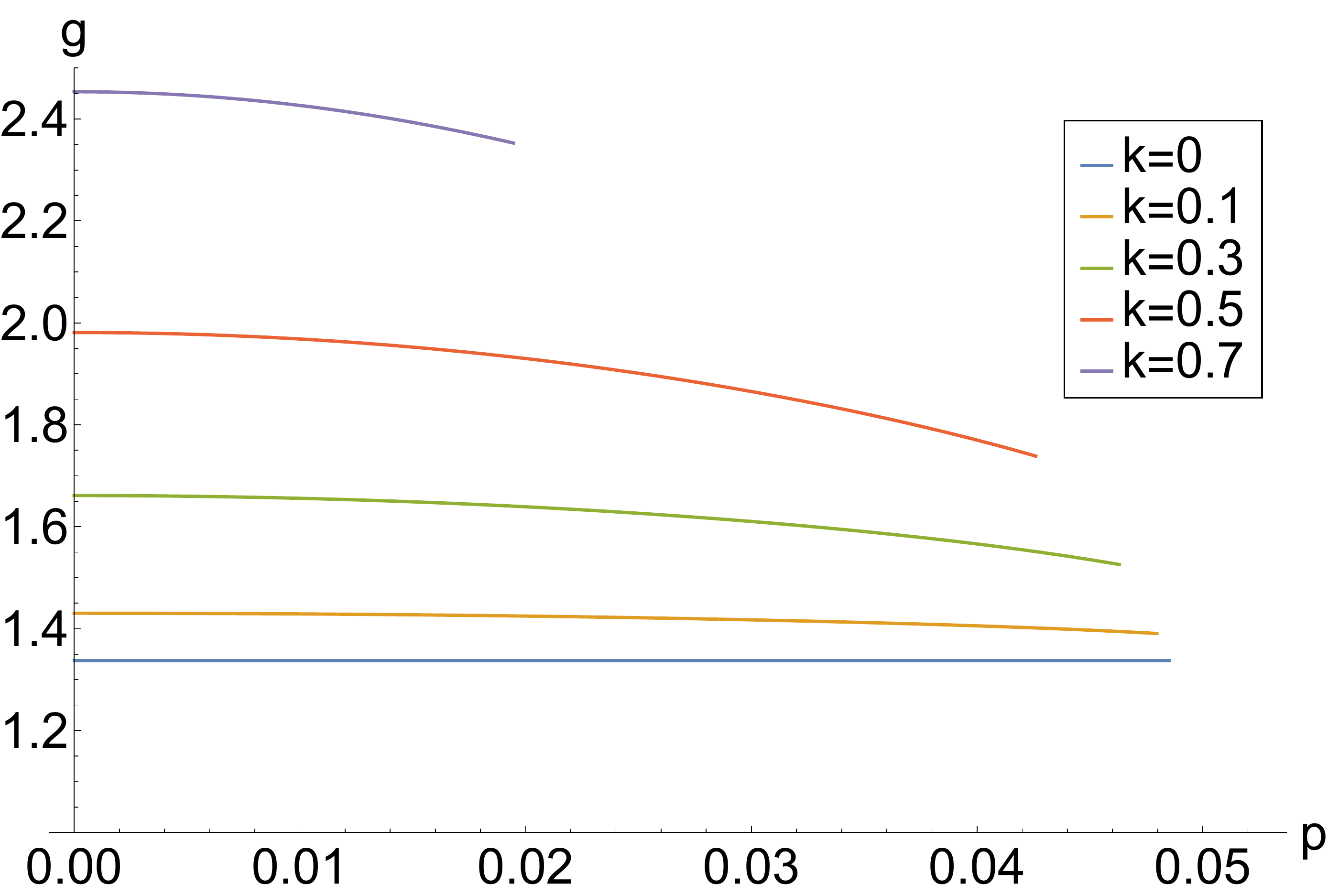}
  \caption{The numerical result of $g$ defined in \eqref{eq:zetagammarelation}. The ratio is plotted as a function of $p$ for different $k$. The values of $(k,p)$ are within the region where the conformal solution are considered to be stable, which we will discuss in next subsection and denoted as region A in Fig.~\ref{fig:kpdiagram}.  We note that these numerical results all rely on the large $M$ limit.}\label{fig:aGaS}
\end{figure}

 In Fig.~\ref{fig:aGaS} numerical values $g=(2\alpha_G)/(\pi^2 \alpha_S)$ as a function of $k$ and doping $p$ are plotted. This is done by solving $\theta_f,\theta_b$ from the Luttinger constraint \eqref{eq:Ltf},\eqref{eq:LtB} and then evaluating \eqref{eq:aGaS}. We found that for fixed $k,p$ there are two solutions for $\theta_b$. The larger $\theta_b$ solution is physical because $\theta_{b,\rm{large}}\to\pi/2$ in the $p\to 0$ limit. The $(k,p)$ values are restricted to be inside the stability region plotted in Fig.~\ref{fig:kpdiagram}.

\subsubsection{$h=1$ sector}

  At $h=1$, there are two degenerate eigenvectors with eigenvalue $1$. Below we show that these two eigenvectors correspond to adding bosonic and fermionic charges, and therefore their corrections in the IR can be absorbed into $\theta_f$ and $\theta_b$.

  We show that the charge modes span the eigenspace of $K_G(1)$. The charges are determined by $\theta_f$ and $\theta_b$ through the Luttinger constraints \eqref{eq:Ltf},\eqref{eq:LtB}. The charge modes are variations of the Green's function, which is
\begin{equation}\label{}
  \delta G_a(C_a,\theta_a;\tau)=\frac{\partial G^c_a}{\partial \theta_a}\delta\theta_a+\frac{\partial G^c_a}{\partial C_a}\delta C_a,
\end{equation} where the variations of $G_a$ come from explicit dependence on $\theta_a$ and implicit dependence through the prefactor $C_a$. Evaluating on the explicit form of Green's function \eqref{eq:gft=0}, we have
\begin{equation}\label{eq:deltaGtheta}
  \delta G_a(\tau)=\left(\frac{\delta C_a}{C_a}+\cot(\theta_a+\frac{\pi}{4}\sgn \tau)\delta\theta_a\right)G^c_a(\tau).
\end{equation}
The $\delta C_a$'s can be determined from $\delta\theta_a$'s by differentiating \eqref{eq:CfCb}
\begin{equation}\label{eq:deltaCfCb}
  \begin{split}
     \frac{\delta C_f}{C_f} & = \left(1+\frac{k t^2 C_b^2\cos 2\theta_b}{J^2 C_f^2\cos 2\theta_f}\right)\frac{\tan 2\theta_f}{2}\delta\theta_f-\frac{kt^2C_b^2\sin2\theta_b}{2C_f^2J^2\cos2\theta_f}\delta\theta_b, \\
     \frac{\delta C_b}{C_b} & = \left(1-\frac{k t^2 C_b^2\cos 2\theta_b}{J^2 C_f^2\cos 2\theta_f}\right)\frac{\tan 2\theta_f}{2}\delta\theta_f+\frac{kt^2C_b^2\sin2\theta_b}{2C_f^2J^2\cos2\theta_f}\delta\theta_b.
  \end{split}
\end{equation}

Combining \eqref{eq:deltaGtheta},\eqref{eq:deltaCfCb} and \eqref{eq:CfCb}, direct computation verifies that
\begin{equation}\label{}
  K_G(1)\delta G=\delta G,
\end{equation} which shows that $\delta G$ spans the eigenspace of unit eigenvalue of $K_G(1)$.


\subsection{Stability constraints of the conformal solution}\label{sec:sta}

\begin{figure}[t]
  \centering
  \includegraphics[width=0.5\textwidth]{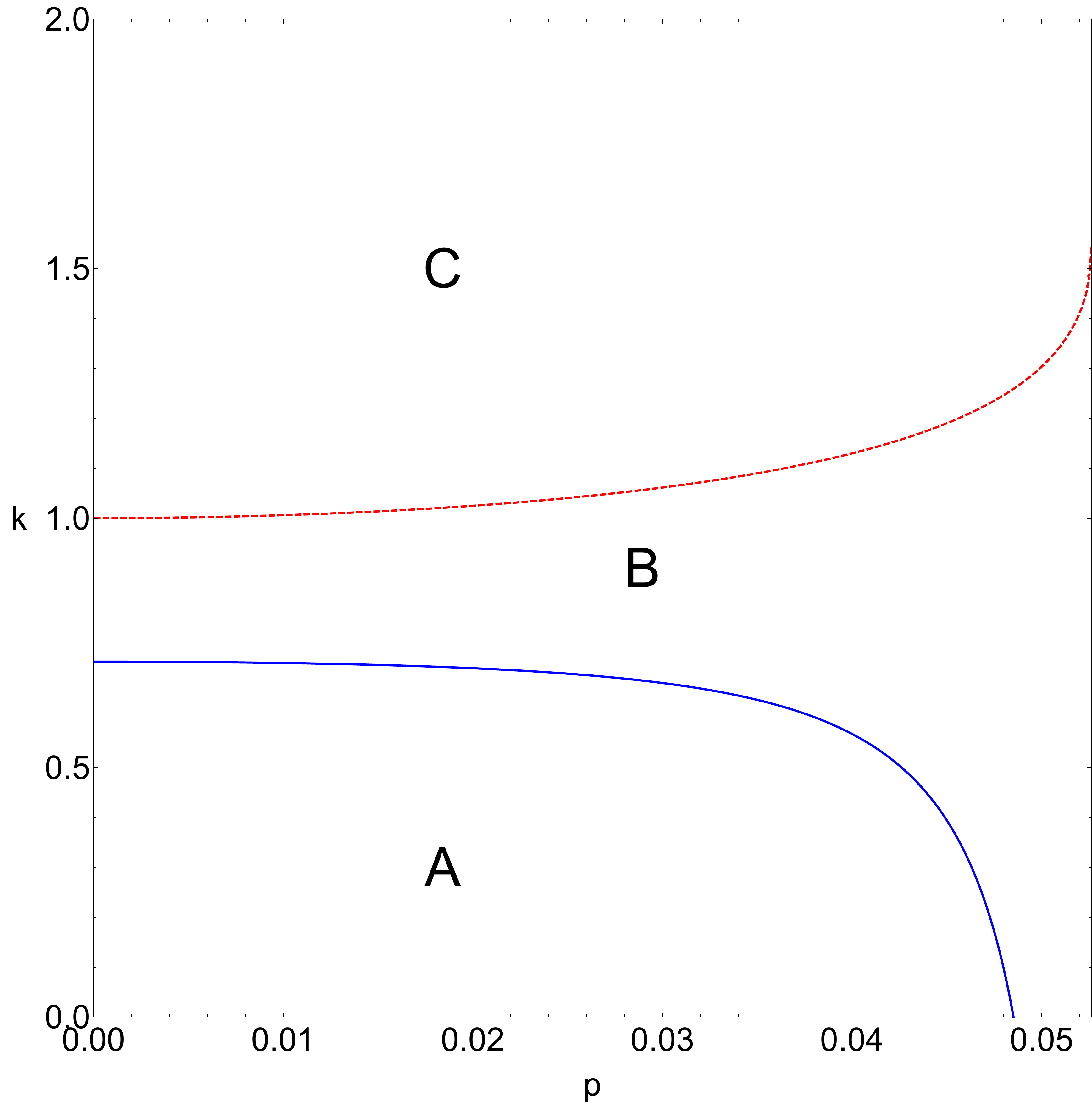}
  \caption{Stable region in $(k,p)$ space in the large $M$ limit. Only region A is stable. The solid line separating region A and B is given by constraint 1. The dashed line separating region B and C is given by constraint 2.}\label{fig:kpdiagram}
\end{figure}

In this subsection we use the resonance theory to analyze stability constraints of the conformal solution \eqref{eq:gft=0} and derive bounds on the $k=M'/M$ and doping density $p$ in the large $M$ limit. We consider the following two criteria for stable.
\begin{enumerate}
  \item The conformal solution should be stable against UV perturbations, i.e. UV sources should be irrelevant. A resonant $h$ for which $K_G(h)$ has eigenvalue one corresponds to an operator $\mathcal{O}_h$ with scaling dimension $h$ in the conformal filed theory. Therefore, $K_G(h)$ should have no resonance within $1/2<h<1$, and $0<h<1/2$ follows from symmetry of $K_G(h)$.
  \item The conformal solution with $\Delta_f=\Delta_b=1/4$ should be stable against the other solution obtained in \cite{Joshi2019} which has $1/4<\Delta_f<1/2$, $\Delta_f+\Delta_b=1/2$. One necessary condition is that the coefficients of Green's functions $C_f,~C_b$ obtained from \eqref{eq:CfCb} should be real.
\end{enumerate}
  The outcome of the above constraints is the diagram Fig.~\ref{fig:kpdiagram} in $(k,p)$ space, where only region A is stable. Region B violates constraint 1 and region C violates constraint 2. Therefore if the conformal solution is stable against UV perturbations, it is automatically stable against the other conformal solution. This agrees with the renormalization group analysis in \cite{Joshi2019} that the $\Delta_f=\Delta_b=1/4$ fixed point is stable and the $1/4<\Delta_f<1/2$ fixed point is a saddle point.
  Therefore the leading UV corrections to Green's function in the stable region indeed come from the $h=2$ resonance.

\begin{figure}
  \centering
  \begin{subfigure}{0.32\textwidth}
    \includegraphics[width=\linewidth]{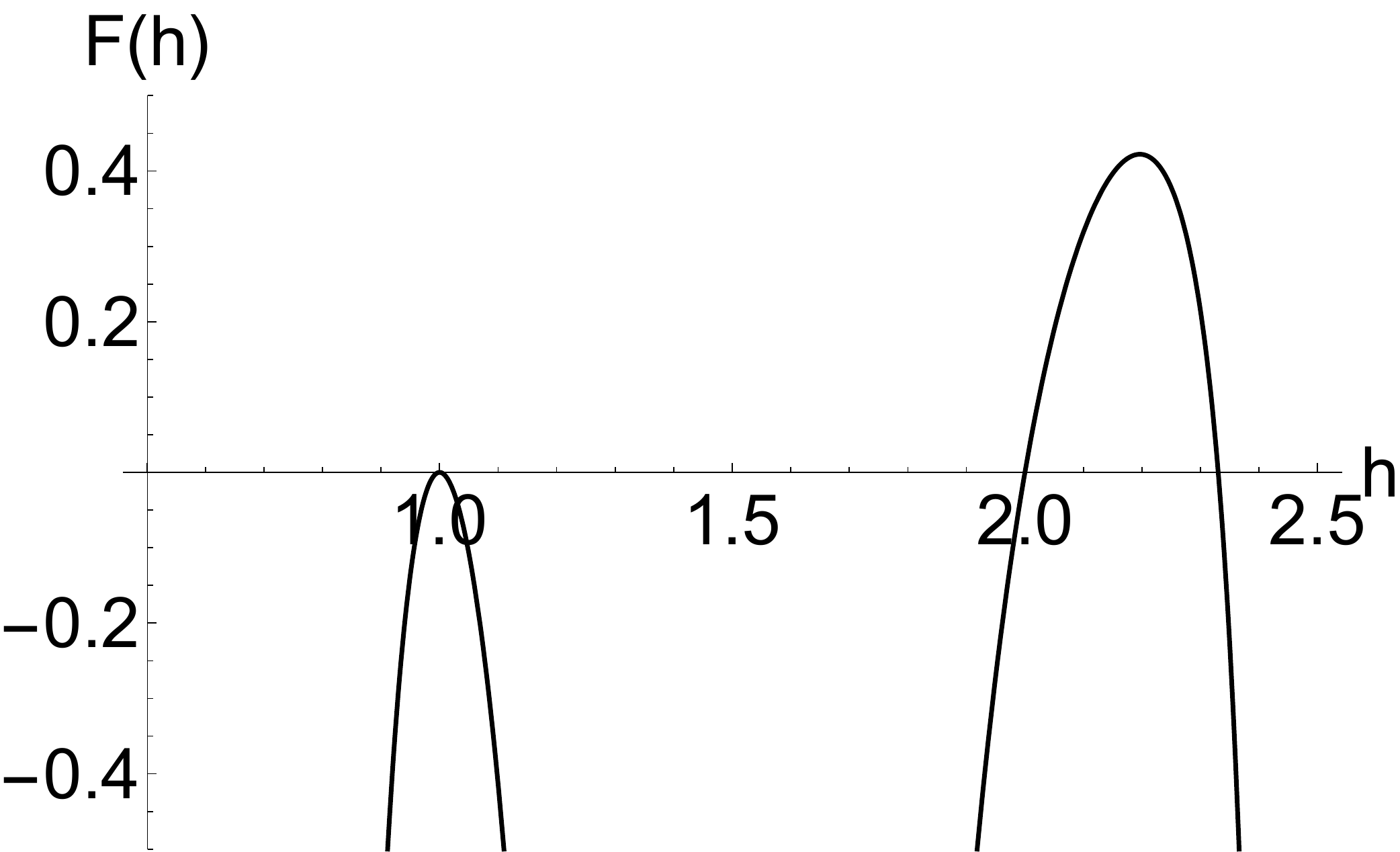}
    \caption{Region A: $F(h)$ at $(k,p)=(0,0)$}
  \end{subfigure}
  \begin{subfigure}{0.32\textwidth}
    \includegraphics[width=\linewidth]{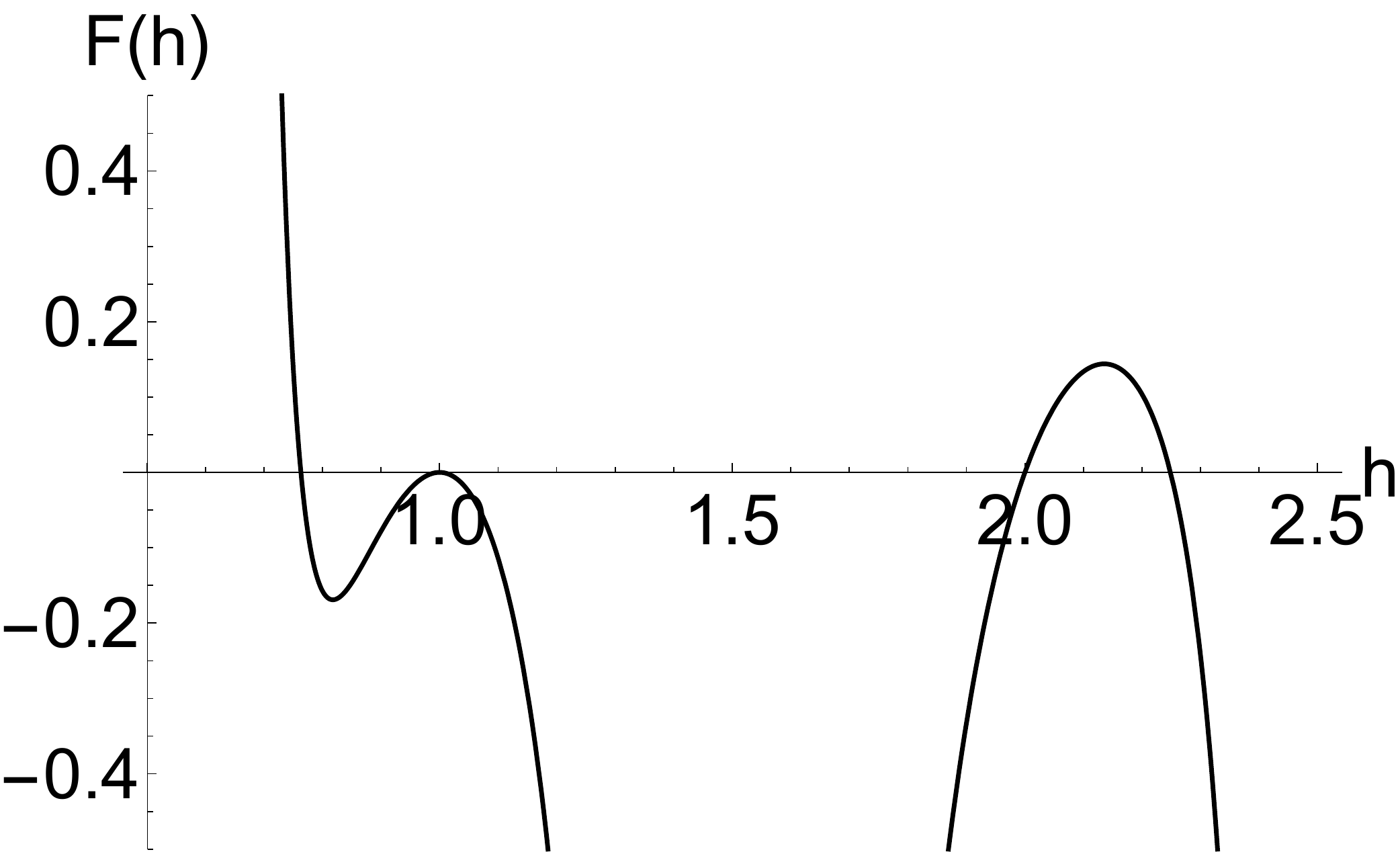}
    \caption{Region B: $F(h)$ at $(k,p)=(0.8,0)$}
  \end{subfigure}
  \begin{subfigure}{0.32\textwidth}
    \includegraphics[width=\linewidth]{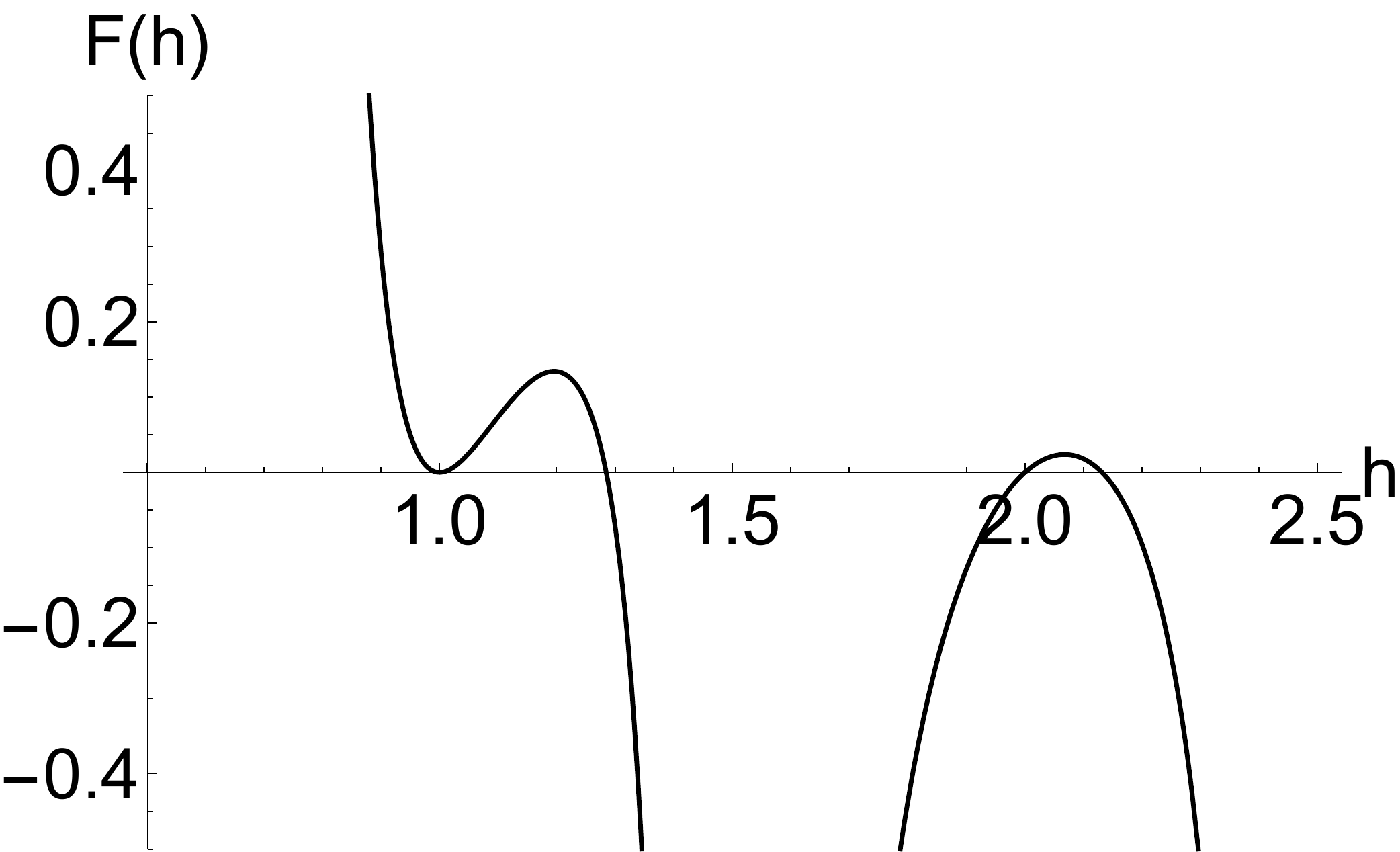}
    \caption{Region C: $F(h)$ at $(k,p)=(1.3,0)$}
  \end{subfigure}
  \caption{Typical behaviors of $F(h)$.}\label{fig:Fplot}
\end{figure}

  Below we will elaborate on how these constraints are implemented. To find resonance, we analyze zeroes of the function
\begin{equation}\label{}
  F(h)=\det(1-K_G(h)),
\end{equation} where the values of $\theta_f,\theta_b$ are determined by $(k,p)$ through the Luttinger constraints \eqref{eq:Ltf},~\eqref{eq:LtB}. There are two solutions for $\theta_b$ and we choose the larger one which has the correct limiting behavior $\theta_b\to\pi/2$ as $p\to 0$. We have also chosen $kp<1/2$ and $p<0.526$ such that the spectral constraint $-\pi\Delta_f<\theta_f<\pi\Delta_f$, $\pi\Delta_b<\theta_b<\pi/2$ is fulfilled. Due to the $h\leftrightarrow 1-h$ symmetry, we focus on the $h>1/2$ case. Typical behaviors of $F(h)$ is shown in Fig.~\ref{fig:Fplot}.

In region A the first two zeroes of $F(h)$ is exactly at $h=1$ and $h=2$ and the conformal solution is stable. When we go from region A to region B, the pole at $h=1/2$ switches sign and a zero develops between $1/2<h<1$. This zero corresponds some relevant operator in the UV source which destabilize the conformal solution. Therefore constraint 1 can be implemented as requiring the pole (4th order) of $F(h)$ at $h=1/2$ should have negative coefficient, which is equivalent to
\begin{equation}\label{eq:cons1}
\begin{split}
 &-8 \left(4 \cos 2 \theta _b+\pi \right) \cos 2 \theta _f-2 (\pi -2) k \cos 2 \theta _b \left(8 \cos 2 \theta _b+3 \pi -4\right) \sec 2 \theta _f-16 k \cos 4 \theta _b\\
 &-4 (3 \pi -4) (k+1) \cos 2 \theta _b-16 k+(4-3 \pi ) \pi>0.
\end{split}
\end{equation} The above condition yields the curve separating region A and region B.

When we further increase $k$ and go to region C, the previous zero in $1/2<h<1$ is now shifted to $1<h<3/2$ ($F(3/2)$ is divergent). This resonance appearing between $1<h<3/2$ can be interpreted as the fermion operator in the $1/2<\Delta_f<1/4$ phase. At the transition point $F''(1)$ switches sign. The condition $F''(1)<0$ is in fact algebraically equivalent to the condition that \eqref{eq:CfCb} has real solutions for $C_f,~C_b$. The latter requires
\begin{equation}\label{eq:cons2}
  1+k\frac{\cos 2\theta_b}{\cos 2\theta_f}>0.
\end{equation} The LHS is actually a factor of $F''(1)$ and all the other factors are negative definite. Therefore \eqref{eq:cons2} implements constraint 2 and gives the boundary between region B and region C.

\section{Finite temperature generalization}\label{sec:finiteT}

  The corrections obtained in the previous section are only valid for zero temperature, and in this section we will generalize them to finite temperature. Because the matrices $W_\Sigma$ and $W_G$ are not only SL(2,$R$) covariant but also reparameterization and $\rm{U(1)}$ covariant, the eigenvector coefficients $\delta G_{a\pm},~\delta \Sigma_{a\pm}$ and the eigenvalues are the same at finite temperature. However, the power law $|\tau|^{1-h}$ in the ansatz \eqref{eq:ansatz} shall be replaced by its counterpart on the thermal circle, which we now calculate.

  In the zero temperature resonance formalism, the eigenfunctions are written in a basis $\tau>0$ and $\tau<0$ branch. It is convenient at zero temperature as conformal transformations doesn't mix positive and negative times. However this is not so at finite temperature because $\pm\tau$ branches have to connect and they are not independent. We instead choose to restrict functions on the interval $0<\theta<2\pi$ ($\theta=2\pi\tau/\beta$ and in this part we take thermal circle size $\beta=2\pi$), and consider symmetric and antisymmetric functions on this interval. Also, this approach applies to both bosons and fermions because they only differ by how the function should be continued beyond $0<\theta<2\pi$.

  First, we need to generalize the Green's function to finite temperature, this can be achieved by a combination of reparameterization and $\rm{U(1)}$ gauge transformation. The result is
\begin{equation}\label{}
\begin{split}
  G_a^c(\theta)&=-\frac{C_a\Gamma(2\Delta_a)}{\pi}\frac{\sin(\theta_a+\pi\Delta_a)}{\left|2\sin\frac{\theta}{2}\right|^{2\Delta_a}}e^{-\ce_a\theta},\quad \theta\in(0,2\pi), 
\end{split}
\end{equation}
where $\ce_a$ is determined by
\begin{equation}\label{eq:epsilon_theta}
  e^{2\pi\ce_a}=\frac{\sin(\theta_a+\pi\Delta_a)}{\sin(\theta_a-\pi\Delta_a)}\zeta_a,\quad e^{-2i\theta_a}=\frac{\sin\pi(i\tilde{\ce}_a+\Delta_a)}{\sin\pi(i\tilde{\ce}_a-\Delta_a)},
\end{equation}
 where $\zeta_b=1,\zeta_f=-1$ indicate the periodicity of $G_a(\theta)$, and $\tilde{\ce}_b=\ce_b,\tilde{\ce}_f=\ce_f+1/2$.

 The corrections $\delta G_a(\theta_1,\theta_2)$ are eigenfunctions of the kernel $K_G=W_\Sigma W_G$. It is argued in \cite{kitaev2017} using star-triangle transformations that these eigenfunctions can be derived from conformal three-point functions
\begin{equation}\label{eq:deltaGa}
  \delta G_a(\theta_1,\theta_2)\propto \int\rd \theta_0 e^{im\theta_0}\langle \chi_a(\theta_1)\chi_a^\dagger(\theta_2)\mathcal{O}_h(\theta_0)\rangle=G_a^c(\theta_{12})f_m(\theta_{12}),
\end{equation}
  where $\mathcal{O}_h$ is an operator of scaling dimension $h$, and $\chi_a$ denotes the fermions and bosons. The resonant response corresponds to $m=0$ and a discrete spectrum of $h$.
 There are two kinds of three-point functions which yield symmetric/antisymmetric corrections under $\theta_{12}\to2\pi-\theta_{12}$:
\begin{eqnarray}
  \langle \chi_a(\theta_1)\chi_a^\dagger(\theta_2)\mathcal{O}_h(\theta_0)\rangle_S &\propto& \frac{1}{|2\sin\frac{\theta_{12}}{2}|^{2\Delta_a}}\frac{|\sin\frac{\theta_{12}}{2}|^h}{|2\sin\frac{\theta_{10}}{2}\sin\frac{\theta_{20}}{2}|^h}, \\
  \langle \chi_a(\theta_1)\chi_a^\dagger(\theta_2)\mathcal{O}_h(\theta_0)\rangle_A &\propto&  \frac{\sgn(\theta_{10})\sgn(\theta_{20})}{|2\sin\frac{\theta_{12}}{2}|^{2\Delta_a}}\frac{|\sin\frac{\theta_{12}}{2}|^h}{|2\sin\frac{\theta_{10}}{2}\sin\frac{\theta_{20}}{2}|^h}.
\end{eqnarray}
 Here we have neglected prefactors and the spectral asymmetry $\ce$, because these factors are absorbed into $G_a(\theta_{12})$ in \eqref{eq:deltaGa}. The $1/|2\sin\frac{\theta_{12}}{2}|^{2\Delta_a}$ factor will also be absorbed into $G_a(\theta_{12})$, and therefore the following terms contribute to $f_m(\theta_{12})$:
\begin{eqnarray}
  W^S_{h,m}(\theta_{12}) &=& \int_0^{2\pi}\rd\theta_0 e^{im\theta_0} \frac{|\sin\frac{\theta_{12}}{2}|^h}{|2\sin\frac{\theta_{10}}{2}\sin\frac{\theta_{20}}{2}|^h}, \\
  W^A_{h,m}(\theta_{12}) &=& \int_0^{2\pi}\rd\theta_0 e^{im\theta_0} \frac{|\sin\frac{\theta_{12}}{2}|^h}{|2\sin\frac{\theta_{10}}{2}\sin\frac{\theta_{20}}{2}|^h}\sgn(\theta_{10})\sgn(\theta_{20}).
\end{eqnarray} 

 The symmetric function $W^S_{h,m}$ has been studied in \cite{kitaev2017}, and we reproduce the analysis here. To evaluate the integral, we switch to complex variable $z=e^{i\theta_0}$, and analytical continue
 $$
   |2\sin\frac{\theta_{10}}{2}|^{-h}=|2(1-\cos(\theta_{10}))|^{-h/2}=|2-z e^{-i\theta_1}-z^{-1}e^{i\theta_1}|^{-h/2}.
 $$
 After some manipulations, we obtain
\begin{equation}\label{eq:WA}
  W^S_{h,m}=e^{im\frac{\theta_1+\theta_2}{2}}|2\sin\frac{\theta_{12}}{2}|^h \oint_{|z|=1}\frac{\rd z}{2\pi i}\frac{z^{m+h-1}}{(z-v)^h(z-v^{-1})^h},
\end{equation}where $v=e^{i\theta_{12}/2}$.
\tikzstyle{branchcut}=[thick, red, decorate, decoration={snake, amplitude=0.5}]
\begin{figure}[t]
  \centering
  \begin{subfigure}{0.45\textwidth}
  \centering
  \begin{tikzpicture}
    \draw[blue, near arrow, line width=1] (0,0) circle (2);
    \node at (0,0) {$\times$};
    \node at (1,{sqrt(3)}) {$\times$};
    \node at (1,{-sqrt(3)}) {$\times$};
  \end{tikzpicture}
  \end{subfigure}
  \begin{subfigure}{0.45\textwidth}
  \centering
  \begin{tikzpicture}
    \draw[blue, near arrow, line width=1] (0,0) circle (2);
    \node at (0,0.15) {$\times$};
    \node at (0,-0.15) {$\times$};
    \node at (0.85,{sqrt(3)}) {$\times$};
    \node at (1.15,{sqrt(3)}) {$\times$};
    \node at (0.85,{-sqrt(3)}) {$\times$};
    \node at (1.15,{-sqrt(3)}) {$\times$};
    \draw[branchcut] (0,0.15)--(0.85,{sqrt(3)});
    \draw[branchcut] (0,-0.15)--(0.85,{-sqrt(3)});
    \draw[branchcut] (1.15,{sqrt(3)})--(3,{sqrt(3)});
    \draw[branchcut] (1.15,{-sqrt(3)})--(3,{-sqrt(3)});
  \end{tikzpicture}
  \end{subfigure}
  \caption{Left: The unregulated integral. The blue line indicates integration contour. Right: The regulated integral. The crosses indicate poles or branch points. The red wavy lines indicate branch cuts.}\label{Fig:contour}
\end{figure}
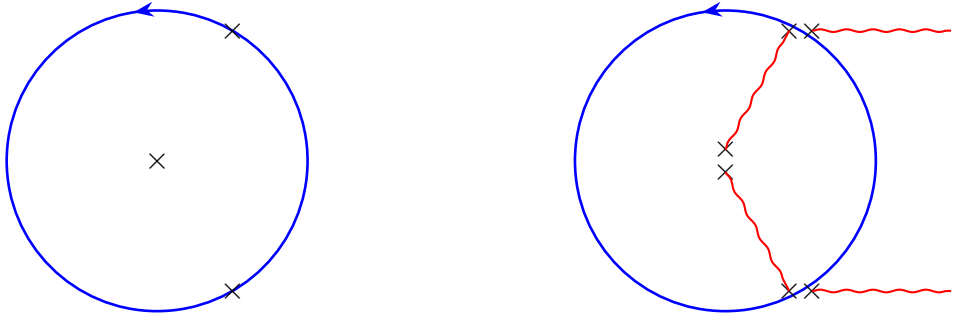
The integrand above has branch points at $0,v,v^{-1},\infty$. We need to regulate the integral in a symmetrical manner with respect to $v,v^{-1}$. To do so, we choose to replace $\sin^2\frac{\theta_{10}}{2}\to \sin^2\frac{\theta_{10}}{2}+\eta^2$, which amounts to splitting each branch point into two, and then we can choose a symmetrical branch cut as shown in Fig.~\ref{Fig:contour}. After that, we deform the integration contour into two loops that winds around the two branch cuts. Along the branch cuts respectively, the integral \eqref{eq:WA} becomes the Euler integral representation of hypergeometric functions, and we obtain 
\begin{equation}\label{}
   W^S_{h,m}(\theta_1,\theta_2)=\frac{e^{im(\theta_1+\theta_2)/2}}{2\cos\frac{\pi h}{2}}\frac{\Gamma(h+m)}{\Gamma(h)}C^S_{h,m}(e^{i\theta_{12}}),\quad0<\theta_{12}<2\pi,
 \end{equation}
 \begin{equation}\label{}
   C^S_{h,m}(u)=A^+_{h,m}(u)+A^-_{h,m}(u^{-1}),\quad A_{h,m}(u)=u^{m/2}(1-u)^h\mathbf{F}(h,h+m,1+m;u).
 \end{equation} Here the $\pm$ in $A^\pm_{h,m}(u)$ means we analytically continue from $0<u<1$ through upper/lower half-plane. 
The calculation of $W_{h,m}^A$ is similar. The difference is that the integrand switches sign when we $z$ passes $v$ or $v^{-1}$, so the prefactors of hypergeometric functions change. The result is
 \begin{equation}\label{}
   W^A_{h,m}(\theta_1,\theta_2)=\frac{e^{im(\theta_1+\theta_2)/2}}{2\sin\frac{\pi h}{2}}\frac{\Gamma(h+m)}{\Gamma(h)}C^A_{h,m}(e^{i\theta_{12}}),\quad0<\theta_{12}<2\pi,
 \end{equation}
 \begin{equation}\label{}
   C^A_{h,m}(u)=iA^+_{h,m}(u)-iA^-_{h,m}(u^{-1}),\quad A_{h,m}(u)=u^{m/2}(1-u)^h\mathbf{F}(h,h+m,1+m;u).
 \end{equation}
To extract the $T=0$ asymptotics, we need to take the $\theta_{12}\to 0$ limit. This is more transparent after applying transformation formulae of hypergeometric functions to $W^A$ and $W^S$:
\begin{eqnarray}
  W^S_{h,m} &=& \frac{e^{im(\theta_1+\theta_2)/2}}{2\cos\pi h}\left(\frac{\Gamma(h+m)}{\Gamma(1-h+m)}e^{i\pi h/2}B_{h,m}^+(e^{i\theta_{12}}) \right. \nonumber \\
  &~&~~~~~~~~~~~~~~~~~~~~ - \left. \frac{\Gamma(1-h)\tan\frac{\pi h}{2}}{\Gamma(h)}e^{i\pi(1-h)/2}B^+_{1-h,m}(e^{i\theta_{12}})\right), \\
  W^A_{h,m} &=& \frac{e^{im(\theta_1+\theta_2)/2}}{2\cos\pi h}\left(\frac{\Gamma(h+m)}{\Gamma(1-h+m)}e^{i\pi h/2}B_{h,m}^+(e^{i\theta_{12}}) \right. \nonumber \\
  &~&~~~~~~~~~~~~~~~~~~~~ \left. -\frac{\Gamma(1-h)\cot\frac{\pi h}{2}}{\Gamma(h)}e^{i\pi(1-h)/2}B^+_{1-h,m}(e^{i\theta_{12}})\right),
\end{eqnarray} where $B^+_{h,m}(u)=u^{m/2}(1-u)^h\mathbf{F}(h,h+m,2h;1-u)$, analytically continued from $0<u<1$ through upper half plane.
  In the above two equations, the first terms can be interpreted as a UV response, and the second terms can be interpreted as IR response. Taking the limit $\theta_{12}\to 0$ for the IR term, we obtain the asymptotics
\begin{eqnarray}
  W^S_{h,m} &\sim& e^{im(\theta_1+\theta_2)/2}\frac{\tan\frac{\pi h}{2}\Gamma(2h-1)}{\Gamma(h)^2}\theta_{12}^{1-h}, \\
  W^A_{h,m} &\sim& e^{im(\theta_1+\theta_2)/2}\frac{\cot\frac{\pi h}{2}\Gamma(2h-1)}{\Gamma(h)^2}\theta_{12}^{1-h}.
\end{eqnarray}
The above results indicate that we can make the following replacement in the ansatz \eqref{eq:ansatz}:
\begin{eqnarray}
  |\tau|^{1-h} &\to& f_{Sh}(\theta_{12})=\frac{W^S_{h,0}(\theta_{12})\Gamma(h)^2}{\tan\frac{\pi h}{2}\Gamma(2h-1)},\quad(\text{In symmetric function}) \\
  |\tau|^{1-h} &\to& f_{Ah}(\theta_{12})=\frac{W^A_{h,0}(\theta_{12})\Gamma(h)^2}{\cot\frac{\pi h}{2}\Gamma(2h-1)},\quad(\text{In antisymmetric function})
\end{eqnarray}

The case of interest is $h=2$, and we have
\begin{eqnarray}
  f_{S2}(\theta) &=& \frac{1}{\pi}\left(1+\frac{\pi-\theta}{2}\cot(\theta/2)\right) ,\\
  f_{A2}(\theta) &=& \frac{\cot(\theta/2)}{2}.
\end{eqnarray} Here $0<\theta<2\pi$.

As a sanity check, we also look at $h=1$, which yields
\begin{eqnarray}
  f_{S1}(\theta) &=& 1, \\
  f_{A1}(\theta) &=& \frac{\pi-\theta}{\pi}.
\end{eqnarray} As discussed previously $h=1$ perturbation corresponds to adding charge to the system, which will introduce a factor $const.\times e^{-\ce\theta}$ in the Green's function, and this matches the solution of $$\frac{\partial G(\theta,\ce)}{\partial\ce}=G(\theta,\ce)(a f_{S1}(\theta)+bf_{A1}(\theta)).$$

It is unspecified that how $f_{Sh}(\theta)$ and $f_{Ah}(\theta)$ should continue beyond $0<\theta<2\pi$, and we are free to choose periodic or antiperiodic boundary condition by including $\sgn(\theta_{12})$ factors. Noticing that the original $\sgn(\theta)$ function is defined on $-\pi<\theta<\pi$, when we periodically continue it to $0<\theta<2\pi$, it transforms the function it multiplies between periodic and antiperiodic boundary conditions.

Taking periodicity into account, we propose that the Green's function correction should have the following form
\begin{equation}\label{eq:deltaGSA}
  \delta G_a(\theta)=\left(\delta G_{Sa} f_{Sh}(\theta)+\delta G_{Aa}f_{Ah}(\theta)\right)G_a^c(\theta),
\end{equation}
If we assume that the resonance theory described in previous section is reparameterization and $U(1)$ covariant, the eigenvalues $k_G(h)$ and eigenvectors $v_h^{R/L}$ should carry over to the current finite temperature setting. If this is true, we can determine the coefficients $\delta G_{Sa}$ and $\delta G_{Aa}$ from $\delta G_{\pm a}$ by a basis change $\pm\to S/A$. The result for $h=2$ correction is therefore
\begin{eqnarray}
  \delta G_b(\theta) &=& -\frac{\alpha_G}{J}(2 f_{S2}(\theta)-3\sin(2\theta_b) f_{A2}(\theta)) G_b^c(\theta), \label{eq:deltaGb}\\
  \delta G_f(\theta) &=& -\frac{\alpha_G}{J}(2 f_{S2}(\theta)-3\sin(2\theta_f) f_{A2}(\theta)) G_f^c(\theta).  \label{eq:deltaGf}
\end{eqnarray}

For the particle-hole symmetric case relevant to the metal-insulator transition \cite{Grisha2020}, we have $\theta_f=0,~\theta_b=\pi/2$, therefore
\begin{eqnarray}
  \delta G_b(\theta) &=& -2\frac{\alpha_G}{J} f_{S2}(\theta) G_b^c(\theta), \\
  \delta G_f(\theta) &=& -2\frac{\alpha_G}{J} f_{S2}(\theta) G_f^c(\theta).
\end{eqnarray}

\section{Evaluation of the conductivity}\label{sec:eval_conductivity}

    In this section we compute the leading correction to resistivity due to $h=2$ resonance.

  The electron Green's function is
\begin{equation}\label{}
  G_c(\theta)=-G_f^c(\theta)G_b^c(-\theta)=-G_f^c(\theta)G_b^c(2\pi-\theta)=-C \frac{e^{-\ce\theta}}{2\sin(\theta/2)},
\end{equation} where $\ce=\ce_f-\ce_b$ and the prefactor $C$ is
\begin{equation}\label{eq:Cc}
  C=\frac{C_f C_b \Gamma(2\Delta_f)\Gamma(2\Delta_b)}{\pi^2}\sin(\theta_f+\pi\Delta_f)\sin(\theta_b-\pi\Delta_b)=\frac{C_f C_b}{\pi}\sin(\theta_f+\pi/4)\sin(\theta_b-\pi/4).
\end{equation}

The $h=2$ correction to the Green's function is 
\begin{equation}\label{}
\begin{split}
  \delta G_c(\theta)&=G_c(\theta)\left(\frac{\delta G_f(\theta)}{G_f^c(\theta)}+\frac{\delta G_b(2\pi-\theta)}{G_b^c(2\pi-\theta)}\right)\\
  &=-\frac{\alpha_G}{ J}\left(4f_{S2}(\theta)-3(\sin(2\theta_f)-\sin(2\theta_b))f_{A2}(\theta)\right)G_c(\theta).
\end{split}
\end{equation} Here in the second line we used \eqref{eq:deltaGb} and \eqref{eq:deltaGf}.

Here are some useful integrals for evaluating the Fourier transforms of $G_c$ and $\delta G_c$:
\begin{align}
  I_{S0}(\omega,\Delta)&=\int_0^{2\pi}\rd\theta\frac{e^{i\omega\theta}}{(2\sin\frac{\theta}{2})^{2\Delta}} = -ie^{-i\pi\Delta}(e^{2\pi i (\Delta+\omega)}-1)B(\omega+\Delta,1-2\Delta)\label{eq:IS0}, \\
  I_{A2}(\omega,\Delta) &= \int_0^{2\pi}\rd\theta\frac{e^{i\omega\theta}}{(2\sin\frac{\theta}{2})^{2\Delta}}f_{A2}(\theta) 
  =\frac{i \omega}{2\Delta}I_{S0}(\omega,\Delta) \label{eq:IA2}, \\
  I_{S2}(\omega,\Delta) &= \int_0^{2\pi}\rd\theta\frac{e^{i\omega\theta}}{(2\sin\frac{\theta}{2})^{2\Delta}}f_{S2}(\theta) 
  =\frac{1}{\pi}\left(I_{S0}(\omega,\Delta)+(\pi+i\partial_\omega)I_{A2}(\omega,\Delta)\right) \label{eq:IS2}.\\
  J_h(\omega,\Delta) &= \int_0^{2\pi}\rd\theta\frac{e^{i\omega\theta}}{(2\sin\frac{\theta}{2})^{2\Delta}} (1-z)^h \mathbf{F}(h,h,1;z)\label{eq:Jh}\\
                     &=-ie^{-i\pi\Delta} (e^{2\pi i (\Delta+\omega)}-1) \Gamma(\omega+\Delta)\Gamma(h-2\Delta+1)~_3\mathbf{F}_2(h,h,\omega+\Delta;1,\omega+h-\Delta+1;1)\nonumber \\
  I_{Sh}(\omega,\Delta) &= \int_0^{2\pi}\rd\theta\frac{e^{i\omega\theta}}{(2\sin\frac{\theta}{2})^{2\Delta}}f_{Sh}(\theta)=\frac{\Gamma(h)^2}{2\sin\frac{\pi h}{2}\Gamma(2h-1)}(J_h(\omega,\Delta)+J_h(-\omega,\Delta)^*) \label{eq:ISh}\\
  I_{Ah}(\omega,\Delta) &= \int_0^{2\pi}\rd\theta\frac{e^{i\omega\theta}}{(2\sin\frac{\theta}{2})^{2\Delta}}f_{Ah}(\theta)=\frac{\Gamma(h)^2}{2\cos\frac{\pi h}{2}\Gamma(2h-1)}(iJ_h(\omega,\Delta)-iJ_h(-\omega,\Delta)^*) \label{eq:IAh}
\end{align}
  Here \eqref{eq:IS0} and \eqref{eq:Jh} are derived using contour integral along the unit circle of $z=e^{i\theta}$, and the cut is from $z=0$ to $z=+\infty$. The integration contour can be deformed to be $\int_1^{(+0)}$, which becomes standard integral representation of beta function and generalized hypergeometric function. \eqref{eq:IA2} and \eqref{eq:IS2} are derived from \eqref{eq:IS0} by integration by parts. \eqref{eq:ISh} and \eqref{eq:IAh} follows from \eqref{eq:Jh} by definition. For $h=2$ correction we will be using Eqs.\eqref{eq:IS0}-\eqref{eq:IS2}.

  In presence spectral asymmetry we should put $\omega=\omega_n+i\ce$ with $\omega_n=n+1/2$ in Eqs.\eqref{eq:IS0}-\eqref{eq:IS2}. The electron Green's function corresponds to $\Delta=\Delta_f+\Delta_b=1/2$, which has log-divergence at the UV. We regulate the divergence using $1/2+\epsilon$ expansion:
\begin{eqnarray}
  I_{S0}(\omega_n+i\ce,1/2+\epsilon) &=& \frac{e^{-2\pi\ce}-1}{2\epsilon}+e^{-\pi\ce}(i\pi\cosh(\pi\ce)-2H(\omega_n-\frac{1}{2}+i\ce)\sinh(\pi\ce)), \\
  I_{A2}(\omega_n+i\ce,1/2+\epsilon) &=& \frac{i\omega_n-\ce}{1+2\epsilon}I_{S0}(\omega_n+i\ce,1/2+\epsilon),
\end{eqnarray}
\begin{align}\label{}
  &I_{S2}(\omega_n+i\ce,1/2+\epsilon) = \frac{e^{-\pi\ce}\cosh(\pi\ce)(\ce-i\omega_n)}{\epsilon}\nonumber\\
   &+ \frac{e^{-\pi  \ce }}{\pi} \left[2 \pi  (\ce -i \omega_n ) \cosh (\pi  \ce ) \left(-1+H(i \ce +\omega_n -\frac{1}{2})\right)\right.\nonumber\\
   &+\left.\sinh (\pi  \ce ) \left(\pi ^2 (-\omega_n -i \ce )+2 (\omega_n +i \ce ) \psi ^{(1)}\left(i \ce +\omega_n +\frac{1}{2}\right)-2\right)\right],
\end{align}where $H(x)=\gamma+\psi(1+x)$ is the harmonic number, and $\psi^{(n)}(x)$ is $n$-th derivative of digamma function $\psi(x)$.
The poles in $\epsilon$ translates to some log terms in $\omega$ which depends on UV behavior of $G_c(\theta)$.

Let us compute the spectral density of the above integrals $A_\alpha(\omega)=-2\Im I_\alpha(i\omega_n\to \omega+i\delta,1/2+\epsilon)$, which yields
\begin{eqnarray}
  A_{S0}(\omega) &=& -2\pi e^{-\pi\ce}\frac{\cosh(\pi\omega)}{\cosh(\pi(\omega-\ce))}, \\
  A_{A2}(\omega) &=& -2\pi e^{-\pi\ce}\frac{\cosh(\pi\omega)}{\cosh(\pi(\omega-\ce))} (\omega-\ce),\\
  A_{S2}(\omega) &=& -2\pi e^{-\pi\ce}\frac{\cosh(\pi\omega)}{\cosh(\pi(\omega-\ce))} (\omega-\ce)\tanh\pi(\omega-\ce).
\end{eqnarray}
The electron spectral density $A_c(\omega)=-2\Im G_c(\omega+i\delta)$ is (we restored the temperature by replacing $\omega\to \frac{\beta\omega}{2\pi}$)
\begin{equation}\label{}
  A_c(\omega)=-CA_{S0}(\frac{\beta\omega}{2\pi}),
\end{equation}
where $C$ is given in \eqref{eq:Cc}.

The correction due to $h=2$ mode is (we have inserted a $2\pi/\beta$ factor to ensure correct dimensionality)
\begin{equation}\label{}
  \delta A_c(\omega)=C\frac{2\pi\alpha_G}{\beta J}\left[4A_{S2}(\frac{\beta\omega}{2\pi})-3(\sin2\theta_f-\sin2\theta_b)A_{A2}(\frac{\beta\omega}{2\pi})\right].
\end{equation}
The conductivity is 
\begin{equation}\label{}
  \sigma_{DC}=\frac{ M' e^2 t^2 a^{2-d}}{2\pi z}\int \rd\omega (A_c(\omega)+\delta A_c(\omega))^2 \beta n_F(\omega)n_F(-\omega),
\end{equation}
The background term is
\begin{equation}\label{eq:sigma0}
\begin{split}
  &\sigma_0=\frac{ M' e^2 t^2 a^{2-d}}{2\pi z}\int \rd \omega A_c(\omega)^2 \beta n_F(\omega)n_F(-\omega)\\
  &=\frac{ M' e^2 t^2 a^{2-d}}{2\pi z}\times 4\pi^2 C^2 e^{-2\pi\ce}\int_{-\infty}^{\infty} \rd x \frac{\cosh^2(x/2)}{\cosh^2(x/2-\pi\ce)}\frac{1}{(e^x+1)(e^{-x}+1)}\\
  &=2\pi C^2 e^{-2\pi\ce}\frac{ M' e^2 t^2 a^{2-d}}{z}.
\end{split}
\end{equation}
The correction term is
\begin{equation}\label{eq:sigma1sigma0}
\begin{split}
  \sigma_1&=\frac{ M' e^2 t^2 a^{2-d}}{2\pi z}\int \rd \omega 2A_c(\omega)\delta A_c(\omega)\beta n_F(\omega)n_F(-\omega)\\
  &=\sigma_0\left(-16\frac{\alpha_G}{\beta J}\right)\int_{-\infty}^{\infty} \rd x \frac{\cosh^2(x/2)}{\cosh^2(x/2-\pi\ce)}\frac{(x/2-\pi\ce)\tanh(x/2-\pi\ce)}{(e^x+1)(e^{-x}+1)}\\
  &=\sigma_0\left(-8\frac{\alpha_G}{\beta J}\right),
\end{split}
\end{equation}where the $A2$ term in $\delta A_c$ doesn't contribute as its integrand is an odd function of $x/2-\pi\ce$.
The resistivity is therefore
\begin{equation}\label{}
  \rho(T)=\frac{1}{\sigma_0}\left(1+8\alpha_G\frac{T}{J}+\mathcal{O}(T^2)\right),
\end{equation} with slope $\zeta=8\rho(0)\alpha_G/J$ and $\rho(0)=1/\sigma_0$.

We know that the coefficient of the Schwarzian action is related to heat capacity $C=\gamma T$ by \cite{Maldacena2016,Davison17}
\begin{equation}\label{}
  \gamma=4\pi^2 \frac{\alpha_S}{J},
\end{equation} so we have the relation which reduces to \eqref{eq:zetagammarelation}
\begin{equation}\label{eq:zetagamma2}
  \frac{\zeta}{\rho(0)\gamma}=\frac{2}{\pi^{2}}\frac{\alpha_G}{\alpha_S}, 
\end{equation}where the ratio $\alpha_G/\alpha_S$ is given in \eqref{eq:aGaS}. From numerical values of $\alpha_G/\alpha_S$ shown in Fig.~\ref{fig:aGaS}, \eqref{eq:zetagamma2} is positive for all $(k,p)$ consistent with stability constraints discussed in Sec.~\ref{sec:sta}.

\section{Conclusions}
\label{sec:conc}
 
We have discussed a class of deconfined critical points of electronic models on large dimension lattices with random nearest exchange interactions with zero mean \cite{Joshi2019,Grisha2020}. These critical points are solvable in a large $M$ limit in which the spin symmetry is generalized from SU(2) to SU($M$).
They can also be analyzed by renormalization group methods \cite{Joshi2019}, but we did not use that approach here.

The key feature of these critical points is that the critical theory has an emergent time reparameterization symmetry. This symmetry is broken by `dangerously irrelevant' terms in the action, in a condensed matter terminology \cite{senthil1,DQCP05}, to a SL(2,$R$) symmetry of the low $T$ state. Such a softly broken reparameterization mode is described by an effective Schwarzian action. 
One important consequence of this soft mode is that correlators of primary operators have a $T/J$ correction in the form of (\ref{Gcnc}), with $\alpha_G$ related to the co-efficient of the Schwarzian term. These are all features that the models considered here share with the SYK models \cite{kitaev2015talk,Maldacena2016,Bagrets2016,kitaev2017,GKST2019}, and they are expected to apply more broadly to other critical theories with a time reparameterization symmetry. In particular, they are also expected to hold beyond the large $M$ limit employed by us.

Turning from (\ref{Gcnc}) to the resistivity, we consider the generality of the structure in (\ref{rhoT}). We discuss some features in turn:
\begin{itemize}
\item The conformal theory yields a $T$-independent `residual' resistivity $\rho(0)$ as $T \rightarrow 0$. We noted that this was a consequence of $1/\tau$ decay of the electron Green's function {\it i.e.\/} the scaling dimension $\Delta_f + \Delta_b=1/2$ for the electron operator \cite{Fu18,Joshi2019}. This feature is expected to hold at deconfined critical points beyond the large $M$ limit, and was also found in the all-loop renormalization group analysis of Joshi {\it et al.\/} \cite{Joshi2019}.
\item The leading correction to $\rho(0)$ is linear in $T$ as $T \rightarrow 0$. This follows directly from inserting (\ref{Gcnc}) into the Kubo formula, and is also generally true, beyond the large $M$ limit. 
This makes the time reparameterization soft mode a promising and generic mechanism for the ubiquitous low $T$ linear-in-$T$ resistivity in correlated electron compounds. 
\item There is a universal relationship between the co-efficient of the linear resistivity and the product of $\rho(0)$ and the co-efficient of the linear in temperature specific heat.
Unlike the other features just noted, this is expected to only hold in the large $M$ limit. It relies on the momentum independence of $G_c ({\bf k}, \omega)$ in (\ref{Gklocal}), which is not present in the large $d$ but general $M$ expression in (\ref{Gk}). 
More generally, we do expect that the coefficient of the linear-$T$ resistivity will be proportional to $\alpha_S$, but the proportionality constant will be non-universal. 
\item The linear-in-$T$ resistivity is a subleading $T/J$ correction to $\rho(0)$. This is true in our large $M$ limit, but not the case in the $M=2$ numerical study of Cha {\it et al.} \cite{Cha19}. We expect that corrections beyond our large $M$ limit will reduce the value of $\rho(0)$: note that $\rho(0)$ vanishes in the alternative large $M$ limit of \ref{app:PG}.
\item The linear power-law of resistivity with temperature is robustly pinned to a scaling dimension associated with the time reparameterization mode. This is in contrast to other mechanisms for non-Fermi liquid transport, where the resistivity has a power-law in temperature related to the scaling dimensions of various model-specific operators \cite{Zhang2017,Balents17,Patel2017,DC18,Guo2019,PatelSS19,Faulkner1043,Patel:2014jfa,Hartnoll:2014gba} which can take diverse values.
\end{itemize}

We conclude by mentioning other contexts in which time reparameterization plays a role:
\begin{itemize}
\item
We recall that the time reparameterization soft mode also plays a key role in the holographic connections of SYK-like models to the quantum gravity theory of black holes with a near-horizon AdS$_2$ geometry \cite{SS10,SS15}: it maps onto quantum fluctuations of the boundary between the near and far horizon regions \cite{kitaev2015talk,nearlyads2,kitaev2017,Moitra:2018jqs,Sachdev:2019bjn,Iliesiu:2020qvm}.
\item
The theory of Fermi surfaces in two spatial dimensions coupled to gapless bosons (order parameters or gauge fields) \cite{SSLee17,Lee:2009epi,MS10,mross,MSSDW} has the important feature that the canonical time derivative terms scale to zero at criticality, like the time derivative terms in (\ref{L}). Consequently, these critical theories also have an emergent time reparameterization symmetry \cite{APSSunpub}, and it would be interesting to explore the consequences.
\end{itemize}

\section*{Acknowledgements}

We thank D.~Chowdhury, A.~Georges, O.~Parcollet, A.~A.~Patel, G.~Tarnopolsky, and M.~Tikhanovskaya for valuable discussions.
This research was supported by the U.S. Department of Energy under Grant DE-SC0019030. Y.G.\ is supported by the Gordon and Betty Moore Foundation EPiQS Initiative through Grant GBMF-4306.

\appendix

\section{Alternative large $M$ limit}
\label{app:PG}

This appendix contrasts our approach from the distinct large $M$ limit of the same model taken by Parcollet and Georges \cite{PG98}. Their procedure does not introduce the orbital index, and (\ref{defc}) is replaced by
\beq
c_{i \alpha} = b_{i}^\dagger f_{i \alpha} \,. \label{defcPG}
\eeq
Consequently, the constraints in (\ref{const}) and (\ref{doping}) are replaced by
\bea
\sum_{\alpha=1}^M f_{i \alpha}^\dagger f_{i \alpha} + b_{i}^\dagger b_{i} &=& \frac{M}{2} \nn \\
 \left\langle b_{i}^\dagger b_{i} \right\rangle  &=& \frac{M}{2} p.
\label{dopingPG}
\eea
In such a large $M$ limit, the boson is always condensed at any non-zero $p$, and does not fluctuate. Consequently we can just replace the boson operator by a c-number $b_i \rightarrow \sqrt{Mp/2}$.
Another important difference, needed because of the boson condensation, is that
the hopping scales differently with $M$: instead of (\ref{t1}) they have
\beq
t_{ij} = \frac{2t}{M\sqrt{z}}\,. \label{t2}
\eeq

With $b$ condensed, the ground state of the large $M$ limit of Ref.~\onlinecite{PG98} is always a Fermi liquid. They also examined the behavior at non-zero $T$, and found that there is a crossover to a non-Fermi liquid above a coherence temperature $T_{\rm coh} \sim (tp)^2/J$. Moreover, they found a linear-in-$T$ resistivity for $T_{\rm coh} < T< J$, but with a mechanism different from that in the present paper; note that their linear-$T$ resistivity does not extend down to $T=0$ at any coupling or density. Their momentum-dependent Green's function replaces
(\ref{Gk}) by
\beq
G^{PG}_c ( {\bf k}, i\omega_n) = \frac{1}{ \displaystyle -\frac{2 \epsilon_{{\bf k}}}{M} + \frac{4 t^2}{M^2}
G^{PG}_c (i\omega_n) + \frac{1}{G^{PG}_c (i\omega_n)}}\,. \label{Gk2}
\eeq
Ref.~\onlinecite{PG98} showed that $G^{PG}_c \sim \mathcal{O}(M)$, and so all three terms in (\ref{Gk2}) have the same scaling in $M$.
Specifically, there saddle point equations are
\bea
G^{PG}_c (\tau) &=& \frac{Mp}{2} G_f^{PG} (\tau) \nonumber \\
G_f^{PG} (i\omega_n) &=& \frac{1}{i\omega_n + \mu - (tp)^2 G_f^{PG} (i\omega_n) - \Sigma_f^{PG} (i\omega_n)} \nonumber \\
  \Sigma_f^{PG}(\tau) &=& -J^2 G_f^{PG} (\tau)^2G_f^{PG} (-\tau)
\,.
\label{Gk3}
\eea
These equations should be compared with our saddle point equations \eqref{Eq:EoM5} and \eqref{Eq:EoM1}-\eqref{Eq:EoM4}: the  non-Fermi liquid behavior arises entirely from $\Sigma_f^{PG}$, and there is no $\Sigma_b^{PG}$.
Combining (\ref{Gk2}) and (\ref{Gk3}), we obtain their result
\beq
G^{PG}_c ( {\bf k}, i\omega_n) = \frac{Mp/2}{i \omega_n + \mu - p \epsilon_{{\bf k}} - \Sigma_f^{PG} (i\omega)}
\eeq
which shows that there is no net electron self-energy from the $t$ term, apart from a renormalization of the hopping from $\epsilon_{\bf k}$ to $p \epsilon_{\bf k}$; this feature is quite different from our large $M$ limit. The absence of an electron self-energy from $t$ leads to a vanishing residual resistivity in the large $M$ limit of this appendix \cite{PG98}.

These forms for the electron Green's function and self energy are quite similar to those of the lattices of SYK islands \cite{Zhang2017,Balents17,Patel2017,DC18,Guo2019,PatelSS19}, and hence the similarities of their transport properties to those of Ref.~\cite{PG98}.

\bibliography{syk}

\end{document}